\documentclass[fleqn,usenatbib]{mnras}
\usepackage{natbib}
\usepackage{amsmath}

\newcommand{\logg}{\ensuremath{\log g}}

\newcommand{\feh}{\ensuremath{\protect\rm [Fe/H] }}
\newcommand{\teff}{T$_{\rm eff}$ }
\newcommand{\pris}{\emph{Pristine }}

\usepackage{graphicx}
\usepackage{tablefootnote}

\usepackage{newtxtext,newtxmath}

\usepackage[T1]{fontenc}
\usepackage{ae,aecompl}

\usepackage{graphicx}	
\usepackage{amsmath}	
\usepackage{amssymb}	

\title[The MDF of the MW halo]{The \pris Survey -- VIII. The metallicity distribution function of the Milky Way halo down to the extremely metal-poor regime}

\author[K. Youakim et al.]{K. Youakim$^{1}$\thanks{E-mail: kyouakim@aip.de},
E. Starkenburg$^{1}$,
N. F. Martin$^{2}$,
G. Matijevi\v c$^{1}$,
D. S. Aguado$^{3}$,
\newauthor C. Allende Prieto$^{4,5}$,
A. Arentsen$^{1}$,
P. Bonifacio$^{6}$,
R. G. Carlberg$^{7}$,
J. I. Gonz\'alez 
\newauthor Hern\'andez$^{4,5}$,
V. Hill$^{8}$,
G. Kordopatis$^{8}$,
C. Lardo$^{9}$,
J. F. Navarro$^{10}$,
P. Jablonka$^{6,9}$,
\newauthor R. S\'anchez Janssen$^{11}$,
F. Sestito$^{1,2}$,
G. F. Thomas$^{12}$
\\
$^{1}$Leibniz-Institut f\"ur Astrophysik Potsdam, An der Sternwarte 16, Potsdam 14482, Germany\\
$^{2}$Observatoire astronomique de Strasbourg, Universit\'e de Strasbourg, 11 rue de l'Universit\'e 67000, Strasbourg, France\\
$^{3}$Institute of Astronomy, University of Cambridge, Madingley Road, Cambridge CB3 0HA, UK \\
$^{4}$Instituto de Astrof\'{\i}sica de Canarias,
              V\'{\i}a L\'actea, 38205 La Laguna, Tenerife, Spain\\
$^{5}$Universidad de La Laguna, Departamento de Astrof\'{\i}sica, 
             38206 La Laguna, Tenerife, Spain \\
$^{6}$GEPI, Observatoire de Paris, Universit\'e PSL, CNRS, Place Jules Janssen, 92190, Meudon, France \\
$^{7}$ Department of Astronomy \& Astrophysics, University of Toronto, Toronto, ON M5S 3H4, Canada \\
$^{8}$Universit\'e C\^ote d'Azur, Observatoire de la C\^ote d'Azur, CNRS, Lagrange, Bd de l'Observatoire, CS34229, F-06304 Nice, France\\
$^{9}$Institute of Physics, Laboratoire d'astrophysique, Ecole Polytechnique F\'ed\'erale de Lausanne (EPFL), Observatoire, 1290 Versoix, Switzerland\\
$^{10}$Department of Physics and Astronomy, University of Victoria, P.O. Box 3055, STN CSC, Victoria BC V8W 3P6, Canada\\
$^{11}$UK Astronomy Technology Centre, Royal Observatory Edinburgh, Blackford Hill, Edinburgh, EH9 3HJ, UK\\
$^{12}$NRC Herzberg Astronomy and Astrophysics, 5071 West Saanich Road, Victoria, BC, V9E 2E7, Canada\\  
}
\date{Accepted XXX. Received YYY; in original form ZZZ}

\pubyear{2019}

\begin{document}
\label{firstpage}
\pagerange{\pageref{firstpage}--\pageref{lastpage}}
\maketitle

\begin{abstract}
The \pris survey uses narrow-band photometry to derive precise metallicities down to the extremely metal-poor regime ($\feh < -3$), and currently consists of over 4 million FGK-type stars over a sky area of $\sim 2~500\, \mathrm{deg}^2$. We focus our analysis on a subsample of $\sim 80~000$ main sequence turnoff stars with heliocentric distances between 6 and 20 kpc, which we take to be a representative sample of the inner halo. The resulting metallicity distribution function (MDF) has a peak at $\feh=-1.6$, and a slope of $\Delta$(LogN)/$\Delta\feh = 1.0 \pm 0.1$ in the metallicity range of $-3.4 < \feh < -2.5$. This agrees well with a simple closed-box chemical enrichment model in this range, but is shallower than previous spectroscopic MDFs presented in the literature, suggesting that there may be a larger proportion of metal-poor stars in the inner halo than previously reported. We identify the Monoceros/TriAnd/ACS/EBS/A13 structure in metallicity space in a low latitude field in the anticenter direction, and also discuss the possibility that the inner halo is dominated by a single, large merger event, but cannot strongly support or refute this idea with the current data. Finally, based on the MDF of field stars, we estimate the number of expected metal-poor globular clusters in the Milky Way halo to be 5.4 for $\feh < -2.5$ and 1.5 for $\feh < -3$, suggesting that the lack of low metallicity globular clusters in the Milky Way is not due simply to statistical undersampling.
\newline
\newline

\end{abstract}




\section{Introduction}

The extended stellar halo of our Galaxy is diffuse and full of stellar substructures, continually being discovered as overdensities and kinematical and/or chemical substructures by on-going large surveys, especially now in the era of the Gaia mission \citep[e.g.,][]{Helmi18,Haywood18,Belokurov18,Malhan18,Koppelman18,Myeong18}. Together they relate a complex history in which the halo is built up through many merger events of smaller systems. While the majority and largest events are thought to have happened in the distant past, some are still continuing in the present day \citep[with the most obvious example being the Sagittarius dwarf galaxy;][]{Ibata94}. It is a complicated task to characterize such a complex structure, especially taking into account our vantage point of observing the halo from within the stellar disc. We know that the halo is highly structured, that its average metallicity is much lower than that of the disc(s) or inner Galaxy and that its stellar population has a very low net rotation \citep[see for reviews on the stellar halo][]{Helmi08,Belokurov13,BlandHawthorn16}. It has further been suggested that the character of the stellar halo is dual in nature, consisting of two broadly overlapping components in density, orbits, and chemistry \citep{Carollo07,Beers12}, but this view has also been challenged \citep{Schoenrich11, Schoenrich14} and is complicated further by the discovery of large substructures in the Gaia DR2 data, standing out in their (partly retrograde) orbits and chemistry in different ways \citep[e.g.,][]{Helmi18,Haywood18,Belokurov18,Myeong19}. Furthermore, simulations of merging galaxy systems have shown that in some cases a significant part of the inner halo can consist of in situ material from the original progenitor \citep[as suggested by][]{Abadi06}, rather than accreted material from merging satellites.  

Robust observational results on the metallicity structure of the Galactic halo can help us understand not only its structure -- as stellar substructures often stand out in metallicity space -- but also the processes involved in its formation. The extremely low-metallicity tail of the metallicity distribution function (MDF), for instance, constrains the pace of chemical evolution at its earliest epochs \citep[for a review on this topic see][]{Frebel15}. Metal-rich halo stars, on the other hand, are more likely to be deposited by larger rather than smaller galaxies (or building blocks). Simulations in a $\Lambda$-CDM framework demonstrate that the median metallicity of a stellar halo is primarily determined by the mass of the most massive building block \citep{Robertson05,Deason16,Oman17,DSouza18a}. An interesting pair of galaxies in this respect are the Milky Way and its close neighbour Andromeda. While the Milky Way has many features in common with our neighbouring galaxy Andromeda, one of the more striking differences is the metallicity of their stellar halos, namely that the Andromeda stellar halo is significantly more metal-rich out to large radii \citep[e.g.,][]{Ibata14}. This could be linked to the presence of stellar substructure on vast scales throughout its large stellar halo and to M32 and its Giant Southern Stream \citep[e.g.,][]{McConnachie18,DSouza18b}. Additionally, a stellar halo mass-metallicity relation is also seen in stellar haloes of galaxies outside the Local Group \citep{Harmsen17}. 

To understand the merging history of the Galaxy in detail, it is thus of clear importance to measure the MDF of the Milky Way stellar halo at different radii and through different lines of sight, minimizing selection biases as much as possible. Such measurements will also help to understand the globular cluster population with respect to the field stars, and constrain chemical evolution in the early Milky Way. Unfortunately, the metal-poor end of the MDF in the Galactic halo remains especially poorly constrained. Several previous studies have used large spectroscopic surveys for this purpose, selecting a halo sample either kinematically to build up a local halo sample, by metallicity to study the metal-poor tail, or by stellar type and magnitude to study the halo at different distances \citep[e.g.,][]{Ryan91,Schoerck09,Allende14}. These different approaches result in very different distributions, especially in the very metal-poor regime. One difficulty is to understand all of the selection biases that affect the spectroscopic survey selection and correct for them appropriately. Especially further out and at lower metallicities, spectroscopic samples quickly become sparser. Additionally, distances are difficult to constrain -- even in the era of the Gaia mission -- as the Galactic halo stretches out far beyond the regime where parallax information is reliable.  

Until recently, observational studies constraining the metal-poor tail of the halo MDF have mostly been done using spectroscopic samples, since photometric metallicity calibrations have been limited to $\feh > -2$ \citep[e.g.][]{Ivezic08}. However, several recent studies have shown that it is possible to provide reliable photometric metallicities using SDSS broad-band photometry down to $\feh = -2.5$, if deeper $u$-band photometry is used \citep{An13,An15,Ibata17,Zuo17}. Subsequently, this metallicity limit has been pushed even further with the implementation of narrow-band filters targeting the Ca II H \& K absorption lines in surveys such as \pris and \emph{SkyMapper} \citep{Keller07,Casagrande19}, which have shown that with this technique it is possible to derive metallicities down to the regime of $\feh < -3$ \citep{Starkenburg17, DaCosta19}.

In this paper, we present MDFs derived using data from the \pris survey \citep{Starkenburg17}. \pris uses a narrow-band filter to obtain photometric metallicities for millions of stars over large regions of the norther Galactic halo. We convolve its results with spectroscopic follow-up information from thousands of stars to remove spurious features that may compromise the photometric metallicity determinations. Because of its photometric nature, the \pris survey is (almost) completely unbiased in its selection of stars, although its metallicity determination is most reliable in the metal-poor regime ($\feh < -1$). We therefore focus our work on the derivation of the MDF at low metallicities for a sample of stars that has been selected to be consistent with halo ages and distances.

We describe the photometric survey, as well as the spectroscopic samples used to rescale the MDF in Section \ref{sec:data}. In Section \ref{sec:halo_sel}, we describe the selection of main sequence turnoff (MSTO) stars and a halo sample, and show the resulting MDFs for different distance ranges. In Section \ref{sec:method}, we apply corrections to the photometric MDF, to account for uncertainties in the photometric metallicities and metallicity biases induced by the colour cut used to select the halo sample. We investigate substructures seen in the MDF as a function of observational direction in Section \ref{sec:mdf}. We discuss the potential remaining biases from the halo sample selection in Section \ref{sec:disc}, and compare the metal-poor halo MDF to those derived in other works. In this section, we also discuss the possibility of the inner halo being dominated by a single, large accretion event, the duality of the stellar halo, and the implications that the halo MDF has for the lack of metal-poor globular clusters (GCs) in the Milky Way. In Section \ref{sec:conc}, we summarize and conclude the paper.  

\section{Description of the data}\label{sec:data}

\subsection{Photometric sample}
\begin{figure}
	\includegraphics[width=\columnwidth]{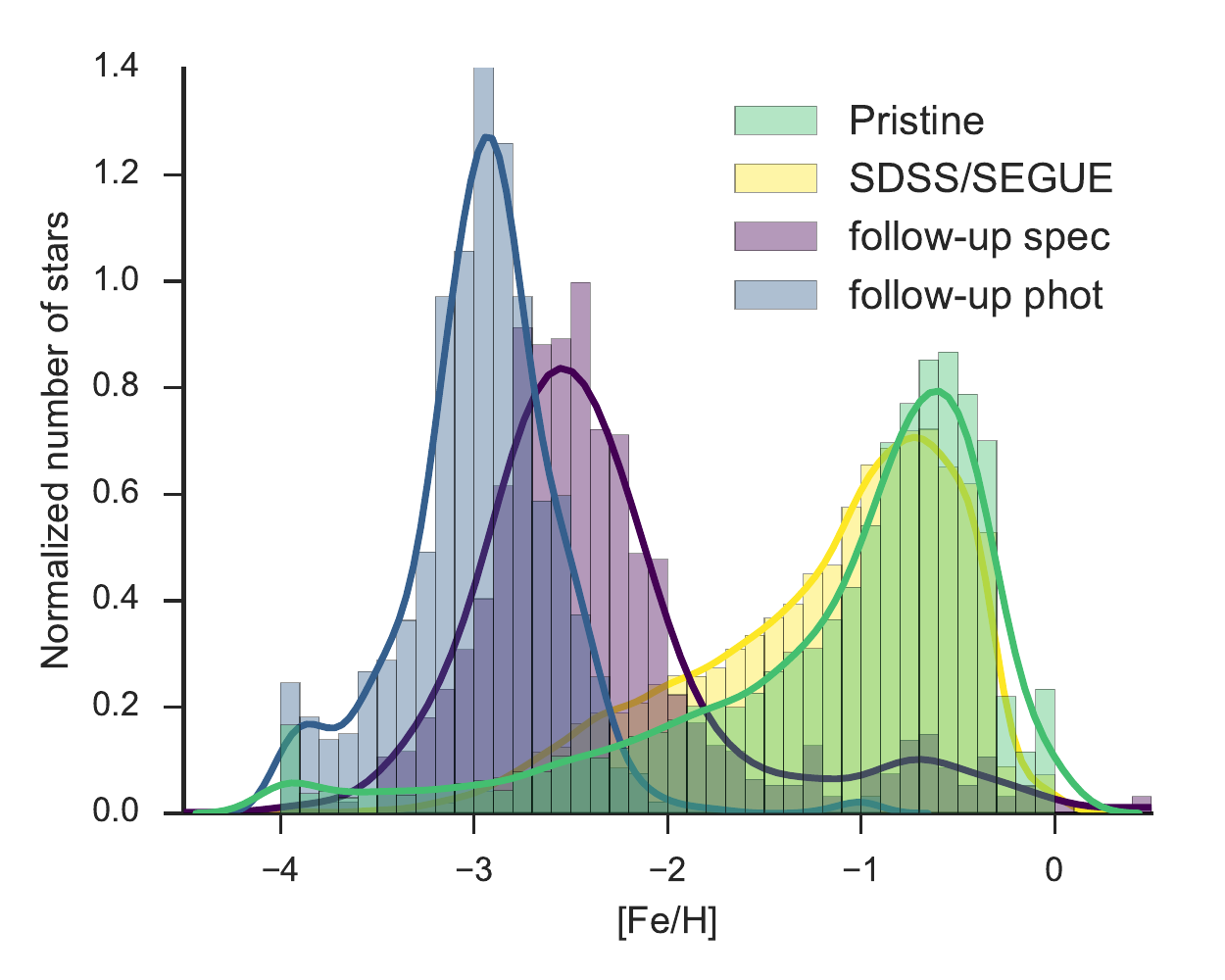}
    \caption{Normalized distributions of metallicities from \pris and the spectroscopic samples. The distribution of the \pris spectroscopic follow-up sample is shown with photometric metallicities (blue) and spectroscopic metallicities (purple), along with the distribution of \pris photometric metallicities for the matched SDSS/SEGUE sample (yellow) and the distribution of the photometric metallicities for the whole \emph{Pristine} sample (green).}
    \label{phot_spec_mdf}
\end{figure}

The \pris survey and the derivation of photometric metallicities are described in detail in \citet{Starkenburg17}. In summary, \pris utilizes a narrow-band filter that is nearly top-hat in shape and centered on the Ca II H \& K absorption lines at $3900-4000$ \AA. These absorption lines are some of the strongest features in a stellar spectrum and their size is directly dependent on the metallicity of the star. The coverage of the survey is $\sim$ 2~500 deg$^2$ in the northern hemisphere and goes to a maximum depth of $g_0 \sim 21.0$ at S/N = 10, although to avoid differential depths of the observed fields and the resulting selection effects, we limit the sample to $g_0 < 20$ in this work. All detected sources from the \pris survey are then matched to stars in the Sloan Digital Sky Survey \citep[SDSS,][]{York00}. The metallicity of each star is determined using a colour-colour space combining the narrow-band information and broad-band information to solve for the temperature of the star (in \citealt{Starkenburg17} and in this work we use SDSS broad-band colours, but a similar principle can be followed using Gaia $G_{BP}$ and $G_{RP}$ colours, as demonstrated in \citealt{Bonifacio19}). Our method is calibrated through the use of overlapping SDSS/SEGUE spectra as well as some stars in the footprint that were additionally observed with high-resolution spectroscopy. Overall, we find that we can determine metallicities for F, G, and K type stars with 4~000 K < \teff\, < 7~000 K, 0.15 < $(g_0 - i_0)$ < 1.5) and $-3.0 < \feh < -0.5$ with $\sim$ 0.2 dex precision \citep{Starkenburg17}. Spectroscopic follow-up has demonstrated that \pris is also successful at identifying stars with $\feh < -3.0$ \citep{Youakim17, Aguado19}. The raw photometric MDF for all stars in the current \textit{Pristine} footprint is shown as the green histogram in Figure \ref{phot_spec_mdf}.

\subsection{Spectroscopic sample}
\label{subsec:spec_samp}
In addition to the photometric component, the \pris survey includes a large spectroscopic follow-up campaign targeting metal-poor star candidates \citep{Youakim17, Caffau17, Bonifacio19, Aguado19, Venn19, Arentsen20}. Although the efficiency of the \pris survey to discover extremely metal-poor stars is very good \citep[for more details see][]{Youakim17, Aguado19} and we take precautions to clean the sample of outliers such as variable stars \citep[using][]{Hernitschek16} and white dwarfs (through additional colour cuts), the metal-poor tail of the MDF is still, to some degree, contaminated by more metal-rich stars. This is to be expected, as the metal-poor population is much less numerous than the more metal-rich stars so that stars of higher metallicity are more likely to scatter down to lower metallicities than the other way around, assuming similar uncertainties. In \citet{Aguado19}, we estimate this contamination to be $\sim$ 18\% for stars that are predicted to be extremely metal-poor ($\feh < -3$) with photometry, but that are found with spectroscopy to be metal-rich ($\feh > -2$). It might also be the case that some unidentified contamination remains, if stars misclassified by photometry are also misclassified by the medium-resolution follow-up spectroscopy, although these cases are likely to be rare. Figure \ref{phot_spec_mdf} shows the raw MDFs, where the blue and purple histograms show the spectroscopic follow-up sample of \pris metal-poor candidates with their predicted photometric metallicities and derived spectroscopic metallicities, respectively. The purple histogram is shifted, showing that the spectroscopic metallicities are more metal-rich than the predicted photometry on average \citep[this effect has been previously described in][]{Starkenburg17}, and also has a tail that extends to high metallicities. For the purpose of building a representative MDF down to the lowest metallicities from photometry, it is important to take these misclassifications into account and correct for them.

\section{Selection of the halo sample}
\label{sec:halo_sel}

The \pris survey is a photometric survey, and as such observes every star in the footprint that is in the right magnitude range (excluding holes caused by chip gaps and stars occluded by bright foreground stars, etc.). However, for each line-of-sight, the total sample will consist of a mixture of stars in different Galactic components, and of different evolutionary stages that probe a large variety of distances at the same magnitude. As it is our goal to produce an MDF of the Galactic halo, we need a way to separate out halo stars from the far more numerous nearby disc stars. This can in principle be done kinematically, chemically, or spatially. In this work, we choose to select a halo sample based primarily on distance from the Galactic plane. The selection of these distance samples and the Galactic environments which they probe are discussed in this section.


\subsection{Distance slices into the Galaxy using MSTO stars}
\label{sec:MSTO}
\begin{figure}
	\includegraphics[width=\columnwidth]{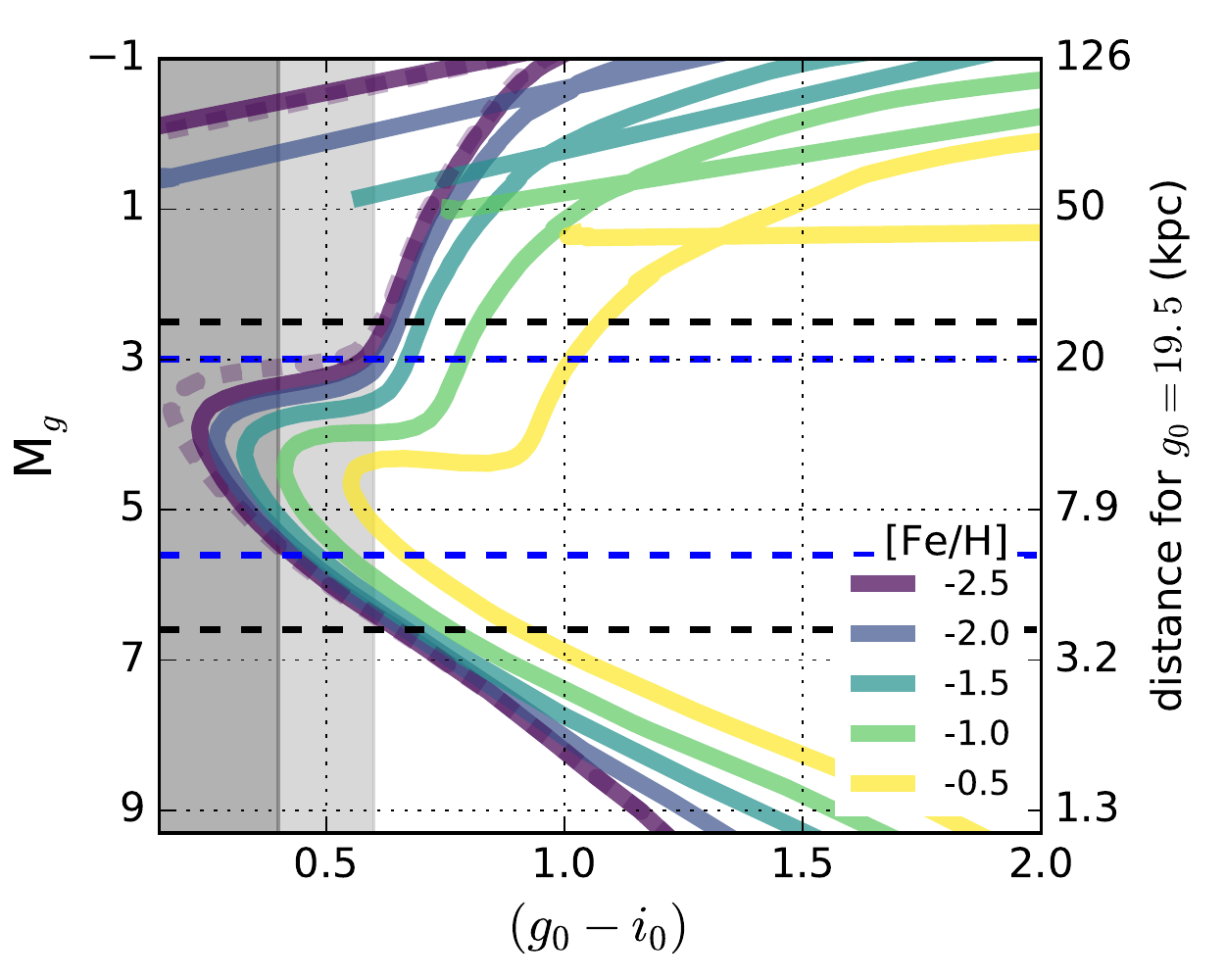}
    \caption{The MSTO colour selection and corresponding distances. PARSEC isochrones of decreasing metallicity and an age of 12 Gyr are plotted as solid lines, and the $\feh = -2.5$ isochrone with an age of 10 Gyr is shown by the dotted line. The absolute SDSS $g$ magnitude is shown on the left axis, and the corresponding distances at an apparent magnitude of $g_0 = 19.5$ on the right axis. The grey shaded region shows the colour cuts used to select the MSTO stars and the dark shaded grey region shows the narrower colour cut used to select the halo sample. The black and blue dashed lines show the corresponding magnitude ranges spanned by the MSTO and halo selections, respectively.}
    \label{CMD_distance}
\end{figure}

 To obtain heliocentric distances for our sample, we make a colour selection to target MSTO stars. Figure \ref{CMD_distance} shows isochrones of different metallicities and a fixed age of 12 Gyr, produced using the PAdova and TRieste Stellar Evolution Code \citep[PARSEC,][]{Bressan12}. We convert [M/H] to \feh\, for these isochrones assuming [$\alpha$/Fe] = +0.4. The dotted purple line shows the isochrone with $\feh = -2.5$ and an age of 10 Gyr, to show how the turnoff shifts as a function of age. We therefore make a colour selection such that we select the full MSTO from the tip of the turnoff up to the sub-giant branch for the most metal-poor isochrone, giving a colour cut of $0.15 < (g_0 - i_0) < 0.6$, shown by the combined dark and light grey regions in Figure \ref{CMD_distance}. The absolute magnitude range spanned by this selection is shown by the black dotted lines, and corresponds to a heliocentric distance range of $\sim4-25$ kpc for a magnitude of $g_0=19.5$ (the middle of the faintest magnitude bin $19 < g_0 < 20$). By selecting different magnitude bins, we separate the Galaxy into different distance ranges, allowing us to target different Galactic environments.

 \subsection{The halo sample}
\label{sec:halo}
 Although the faintest MSTO sample selected with $0.15 < (g_0 - i_0) < 0.6$ and $19 < g_0 < 20$ is likely composed of mostly halo stars, it still probes a wide distance range of $\sim 4-25$ kpc, and thus a wide range of Galactic environments. In order to narrow this range, we impose an additional colour cut at $0.15 < (g_0 - i_0) < 0.4$ (depicted by the dark grey region in Figure \ref{CMD_distance}) to more precisely target the tip of the MSTO. The blue dotted lines show the narrower magnitude range spanned by this sample, which corresponds to a narrower distance range of $\sim6 - 20$ kpc. Therefore, this sample does not probe as far out into the halo as the faintest MSTO sample, but has a larger minimum distance and thus contains less of a contribution from solar neighbourhood and disc stars. 
 
\begin{figure}
	\includegraphics[width=\columnwidth]{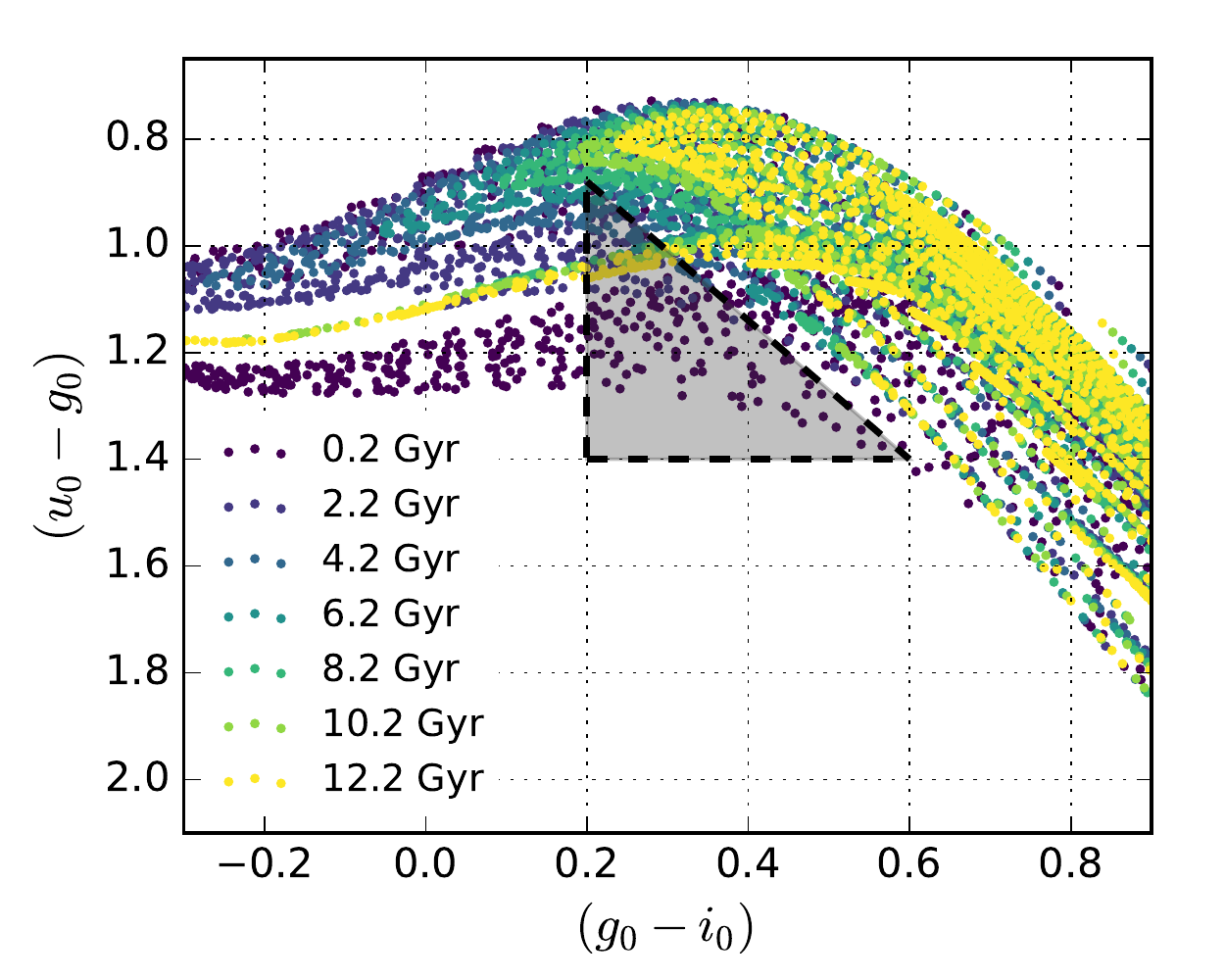}
    \caption{Colour-colour plot of PARSEC isochrones sampling ages of $0.2-12.2$ Gyr and a metallicities of $-2.5 < \feh < -0.5$. The isochrones are coloured by age, but if instead they were coloured by metallicity, the metal-poor populations would be at the top of the plot and the metal-rich ones at the bottom. The grey shaded triangle shows the region removed (see text for details), which targets young disc stars and old horizontal branch stars.}
    \label{u-g_cut_plot}
\end{figure}

 
In a further attempt to clean the sample of young disc stars, we identify a region in the $(u_0-g_0)$ versus $(g_0-i_0)$ diagram that is dominated by young stars. Figure \ref{u-g_cut_plot} shows this colour-colour space with PARSEC isochrones of metallicities ranging from $-2.5 < \feh < -0.5$ and ages from $0.2 - 12.2$ Gyr. The isochrones are colour coded by age, but colour coding them instead by metallicity would show that the more metal-poor isochrones would be at the top of the diagram and the more metal-rich ones would be at the bottom. The grey triangle shows the region that is almost solely populated by young, metal-rich stars. In the upper left of this triangle there are also some older populations, but these are horizontal branch stars, which constitute a much smaller population compared to the more numerous MSTO stars. We therefore decide to remove all stars that fall within this triangular region from the analysis in both the spectroscopic and photometric data sets to clean the young disc dwarfs from the sample. The underlying assumption is that there are far more young stars in the disc than young metal-poor stars in the halo. This cut removes $\sim 10\%$ of the halo sample selected with \pris stars and only $\sim 3\%$ of the spectroscopic sample.


 \begin{figure*}
	\includegraphics[width=\textwidth]{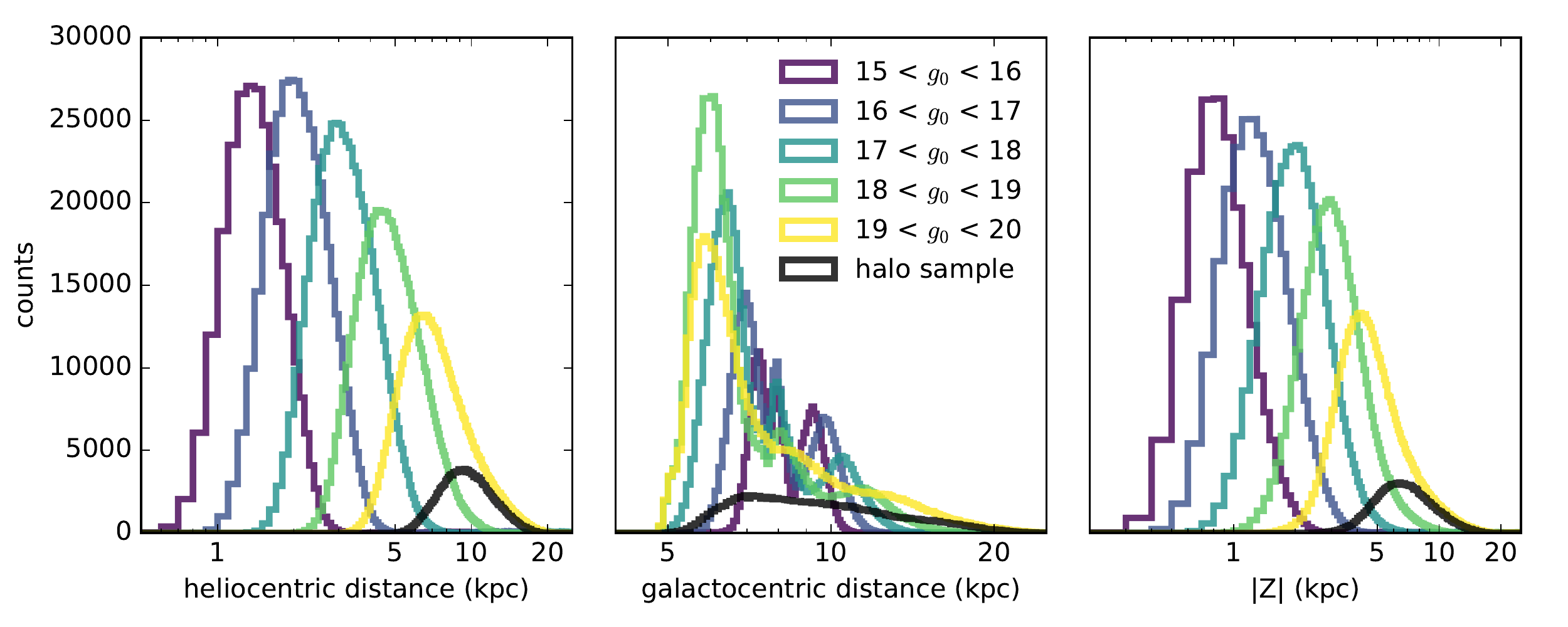}
    \caption{Distance distributions for the MSTO cuts with $0.15 < (g_0 - i_0) < 0.6$ applied to the Galaxia mock catalogues. The left panel shows the heliocentric distances, the middle panel the Galactocentric distances, and the right panel the height above/below the disc, |Z|. Histograms are coloured according to the magnitude range shown, and the black histogram shows the distance distribution for the halo sample with $0.15 < (g_0 - i_0) < 0.4$ and $19 < g_0 < 20$.}
    \label{distance_hist}
\end{figure*}

\subsection{Distance ranges of the samples}
\label{c4sec:distances}


To further investigate the distance ranges probed by these samples, we generate a simulated data set using the Galaxia code \citep{Sharma11}, which implements the analytical Besan\c{c}on model \citep{Robin03} of the Galaxy. We run Galaxia using the default parameters, and selecting the regions of the \pris footprint with a magnitude limit of $14 < g_0 < 22$, an oversampling factor of 3.0, and disc flare and warp turned on. We mimic the survey depth and observational selection function of the \pris survey by selecting a subset of stars from the Galaxia catalogue to match the stellar density of \emph{Pristine}. This is done using SDSS as a reference sample, under the assumption that it is complete at the magnitude limit of the \pris survey ($V < 21$). We then bin the stars spatially (by each 1 x 1 deg$^2$ \pris field), by colour (bins of 0.15 in $(g_0 - i_0)$) and by magnitude (bins of 0.2 in $g_0$) and count the number of stars in each bin. A random subset of Galaxia stars is selected from a given bin, such that the ratio of the subset to the total is equal to the ratio of \pris stars to SDSS stars in that given bin (i.e. $N_{select} / N_{Galaxia} = N_{Pristine} / N_{SDSS}$ for each bin in RA and Dec, colour, and magnitude). Since the \pris sources are cross-matched to SDSS, there are no bins in which $N_{Pristine} > N_{SDSS}$. In addition, the oversampling of Galaxia with fsample = 3 minimizes the cases where $N_{select} > N_{Galaxia}$, and results in very few bins in which this is the case. When this does happen, we simply take all of the Galaxia stars in that bin. The result of this is a Galaxia generated mock sample which has the same coverage as the \pris footprint, and mimics the depth, magnitude and colour distributions of \pris.

\begin{table}
\caption{Heliocentric distance ranges probed by the MSTO samples selected with $0.15 < (g_0 - i_0) < 0.6$ and the shown magnitude cuts. The bottom row shows the heliocentric distance range of the halo sample, selected with $0.15 < (g_0 - i_0) < 0.4$ and $19 < g_0 < 20$. Distance ranges are computed both by using isochrones as well as the distances from the Galaxia mock data. The bold numbers indicate the peaks of the distributions in the left panel of Figure \ref{distance_hist}.}
\centering
\begin{tabular}{ccc}
\hline
Magnitude range    & Isochrone & Galaxia \\ 
& distance (kpc) & distance (kpc) \\ \hline
$15 < g_0 < 16$ & [0.6 - 4]                       & [0.8 - \textbf{1.5} - 3]                       \\ 
$16 < g_0 < 17$ & [1 - 6.3]                         & [1 - \textbf{2} - 4.5]                         \\ 
$17 < g_0 < 18$ & [1.5 - 10]                         & [1.6 - \textbf{3} - 7]                       \\ 
$18 < g_0 < 19$ & [2.4 - 15.9]                      & [2.5 - \textbf{4.5} - 11.2]                        \\ 
$19 < g_0 < 20$ & [3.8 - 25]                        & [3.5 - \textbf{6.5} - 18]                      \\ \hline
halo sample & [6 - 20]                         & [6 - \textbf{9.5} - 16]          \\
\hline
\label{c4:dist_table}
\end{tabular}
\end{table}


We then apply the same selections for the MSTO and halo samples and compute the distance distributions for each in the Galaxia mock catalogue. Figure \ref{distance_hist} shows the heliocentric, Galactocentric, and height above the disc distance distributions for each of the samples, and Table \ref{c4:dist_table} summarizes the specific heliocentric distance ranges probed, computed with isochrones and with the Galaxia catalogue. Both methods yield similar distances, particularly at the lower limit of the distance range, but the isochrone technique has slightly higher upper limits due to the extension of the beginning of the sub-giant branch, which in Galaxia only represents a small number of stars in the tail of the distributions. Using heliocentric distance ranges can sometimes make it difficult to interpret exactly which Galactic environments are being probed, however, as can be seen in the middle and right panels of Figure \ref{distance_hist}, the high Galactic latitudes ($|b| > 20^{\circ}$) and limited depth of the \pris sample result in a distance distribution that still nicely samples the inner Galactic halo when converted to the reference frame of the Galactic center. Most importantly, the halo sample probes a distance range of $\sim 5 - 20$ kpc from the Galactic center, and a minimum height above the disc of $|Z| > 4$ kpc, avoiding in large part the stellar populations of the bulge and disc.

\label{sec:mdf_dist}
\begin{figure}
	\includegraphics[width=\columnwidth]{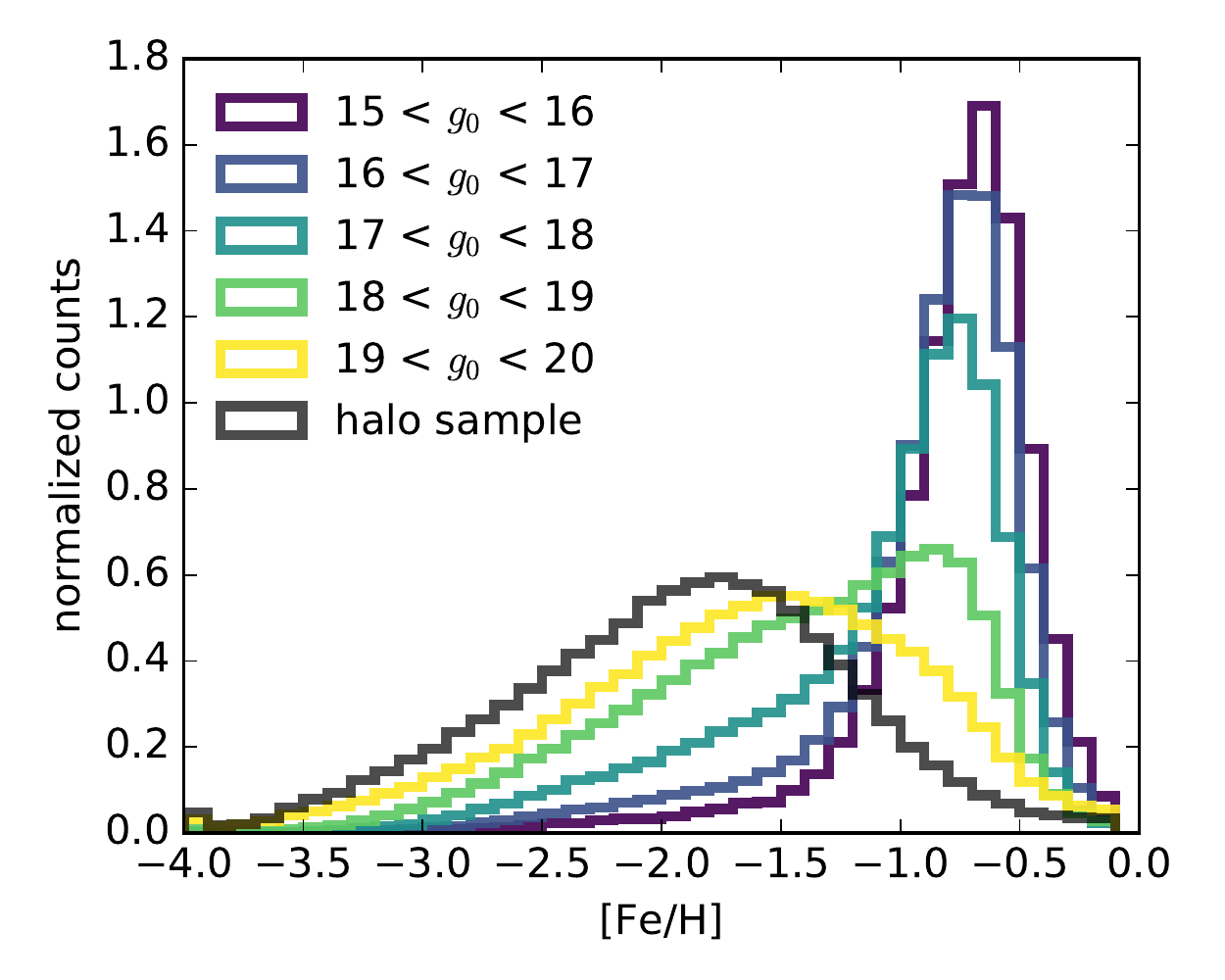}
    \caption{MDFs of MSTO samples selected with a colour cut of $0.15 < (g_0 - i_0) < 0.6$, separated into bins of magnitude. The black histogram shows the halo sample with $0.15 < (g_0 - i_0) < 0.4$ and $19 < g_0 < 20$.}
    \label{mdf_MSTO}
\end{figure}

\subsection{The MDF at different distances}

Figure \ref{mdf_MSTO} shows the resulting MDFs for the MSTO samples at different magnitude ranges. These are raw photometric MDFs from \pris data without any corrections applied, to qualitatively show the decreasing metallicity gradient with increasing heliocentric distance. The coloured histograms are produced with the $0.15 < (g_0 - i_0) < 0.6$ colour cut and limiting the samples to the shown magnitude ranges. The black histogram shows the halo sample with $0.15 < (g_0 - i_0) < 0.4$ and $19 < g_0 < 20$. There is a clear trend that the MDFs shift to lower metallicities with increasing distance. There is also a clear transition from the strong peak at $\feh\, \sim -0.7$ in the brighter samples where the disc dominates, and the metal-poor peak at $\feh\, \sim -1.6$ in the fainter samples where the halo population dominates. The green and yellow histograms in the magnitude range of $18 < g_0 < 20$ show the transition, where both populations are visible.



\begin{figure*}
	\includegraphics[width=\textwidth]{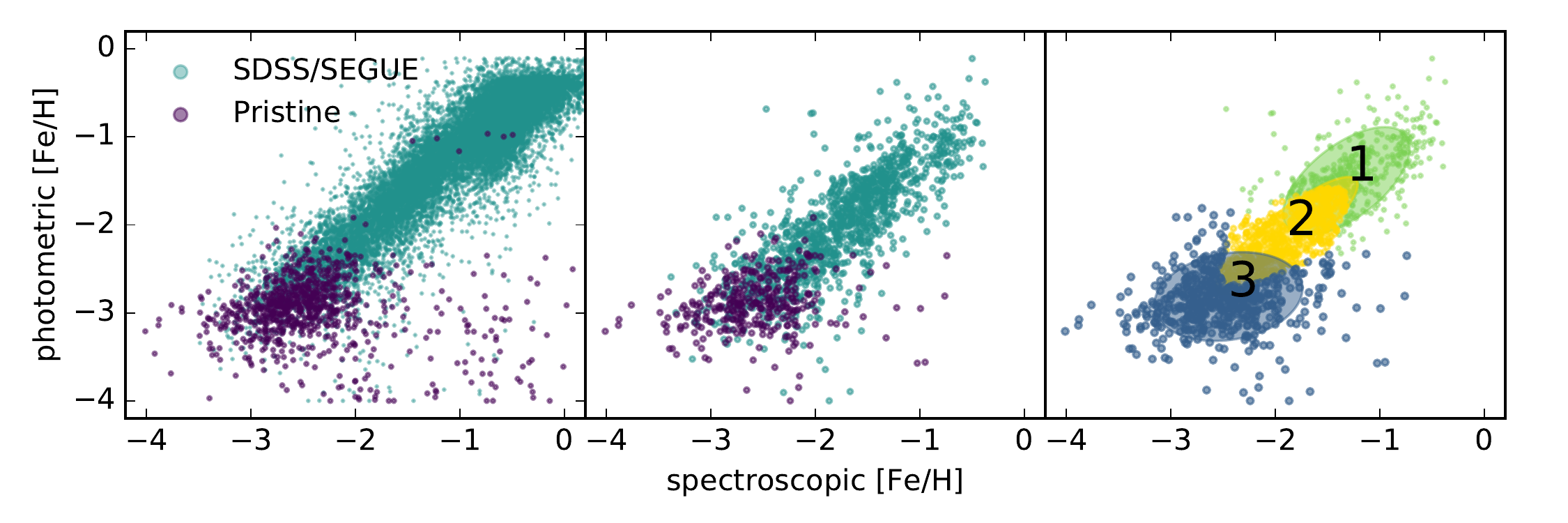}
    \caption{The spectroscopic sample used to rescale the \pris photometric MDF for the halo sample. The left panel shows the full spectroscopic samples from \pris follow-up (purple) and SDSS/SEGUE (teal) and the middle panels shows the samples after applying the colour cut of $0.15 < g_0 < 0.4$ and a magnitude cut of $17 > g_0$ to select the halo sample. The right panel shows the corresponding 3-component GMM used to fit the data. The points are coloured according to which of the Gaussian components they belong, and the 1-$\sigma$ contour of each is shown by the over-plotted oval of the same colour. The varied point sizes are simply to enhance the clarity in crowded regions.}
    \label{gmm}
\end{figure*}

\section{Correcting the MDF}
\label{sec:method}
For the current data set, there are two main corrections that we take into account in order to account for the metallicity biases present in the MDF. The first comes from determining metallicities photometrically and the inherent uncertainties associated with this procedure, particularly in the metal-poor regime, and the second are the biases introduced by making the colour cut of $0.15 < (g_0 - i_0) < 0.4$ to select the distance limited halo sample. The effect of these biases, as well as the methods used to correct them, are discussed in detail in this section.

\subsection{Corrections for the photometric metallicities}
\label{sec:gmm}

Despite cleaning the photometric metallicity sample of contaminating objects such as white dwarfs and variable stars, some degree of contamination still persists, particularly at the metal-poor end. This is mostly due to metal-rich stars preferentially scattering into the metal-poor regime, but can also be caused by stars with unusual colours (e.g. blue stragglers, or stars with long-period variability not picked up in \citealt{Hernitschek16}) or peculiar behaviour in their Ca II H \& K lines (e.g. binaries, or stars with Ca II H \& K in emission). Fortunately, most of these stars that are misclassified photometrically can be identified with the spectroscopic follow-up sample, and can subsequently be corrected for using a statistical rescaling of the photometric MDFs. The left panel of Figure \ref{gmm} shows the relation between the \pris photometric metallicities and the derived spectroscopic metallicities for the spectroscopic sample, which we assume to represent the true metallicity of the star. This sample consists of $\sim$ 1~000 stars from \pris spectroscopic follow-up, which predominantly cover the metallicity range of $-4 < \feh < -2.5$ and are shown by the purple points \citep[see][for more details on this sample]{Aguado19}. To supplement the sample at $\feh > -2.5$, we add $\sim$ 17~000 cross-matched SDSS/SEGUE stars over the \pris footprint that have reliable metallicity determinations from the SEGUE Stellar Parameter Pipeline \citep[SSPP, our selection of this sample is described in detail in][]{Starkenburg17}, shown by the teal points in the Figure. For the 59 stars in common between the two samples, we find no systematic shift between the metallicities derived in the \pris follow-up analysis and the SSPP, and a dispersion of $\sim$ 0.3 dex. The middle panel shows the same samples after the application of the halo selection, with $0.15 < (g_0 - i_0) < 0.4$. We additionally apply a cut of $g_0 > 17$ to the SDSS/SEGUE sample to remove the strong disc feature at $\feh > -1$, and to allow for the distribution to have a larger spread to compensate for the larger uncertainties at faint magnitudes. We do not apply a magnitude cut of $19 < g_0 < 20$ since there are too few stars in the spectroscopic sample at these faint magnitudes. Similarly, we do not make a magnitude cut on the \pris spectroscopic sample, since this sample is restricted to an even brighter magnitude range. Therefore, the implicit assumption that we make is that the relation between photometric and spectroscopic metallicity at $\feh < -2.5$ does not vary significantly between stars with $14 < g_0 < 18$ and $18 < g_0 < 20$. We also apply the $(u_0-g_0)$ vs $(g_0-i_0)$ cut to remove young disc stars as described in Section \ref{sec:halo}, but this only removes $\sim$ 3\% of the stars from the spectroscopic samples. The sizes of the samples after the cuts are applied are 1163 and 340 for the SDSS/SEGUE and \pris samples, respectively.

To perform the rescaling, we build a probability density function (PDF) to describe the relation between the photometric and spectroscopic metallicities. We choose to use a Gaussian mixture model (GMM) for this purpose, because Gaussians are simple functions that are well suited to fitting smooth, continuous data. 


We make the GMM using the python package \textit{sklearn.mixture.GaussianMixture} \citep{scikit_learn}, which implements the expectation-maximization algorithm for fitting distributions composed of several Gaussian components, the number of which are provided as an input parameter. We use the default parameters, with the exception of increasing the maximum number of iterations to 500 to ensure that the solution converges. 




To determine the number of components to use to best fit the data we use the Bayesian information criterion (BIC). We compute GMMs for each of the samples using between $1-10$ components, and then compute the BIC for each using the built-in attribute of the sklearn mixture model package. We repeat this 1~000 times, taking the mean of the resulting distribution of test values, and then choose the number of components with the lowest computed BIC. For the halo sample, this gives a solution with 3-components, and the resulting GMM is shown in the right panel of Figure \ref{gmm}. All three of the components have similar weights, such that the GMM is not dominated by any one component. Table \ref{gmm_params} in the Appendix shows the best fit parameters for each of the Gaussian components, including the means, weights, and covariance matrices. 


To check that the GMM fits the data well, we use a 2D Kolmogorov-Smirnov (KS) test, which begins with the null hypothesis, $H_0$, that two samples are drawn from the same parent distribution. First, we draw a sample from the GMM and compute the KS statistic and p-value between this sample and the original spectroscopic one using a two-sample KS test \footnote{The code used for the 2D KS test was modified from \url{https://github.com/syrte/ndtest}}. We repeat this procedure 1~000 times and take the mean p-value for the set of iterations. We find $p > 0.1$ for a GMM with 3 components, meaning that we cannot reject $H_0$ and the two samples are consistent with being drawn from the same distribution. If we use 2 components, then the resulting p-value is marginal and inconclusive, with $0.1 > p > 0.01$. Using only 1 component gives $0.01 >> p$, and $H_0$ is rejected. Taken together, these tests confirm that the GMM is an appropriate model for fitting a PDF to the spectroscopic sample, as long as the right number of components are used. 

To apply the GMM correction, we take each photometric metallicity from the \pris sample, and simply draw from the corresponding 1-D PDF made by sampling along a horizontal line at that given metallicity in Figure \ref{gmm}.

\begin{figure}
	\includegraphics[width=\columnwidth]{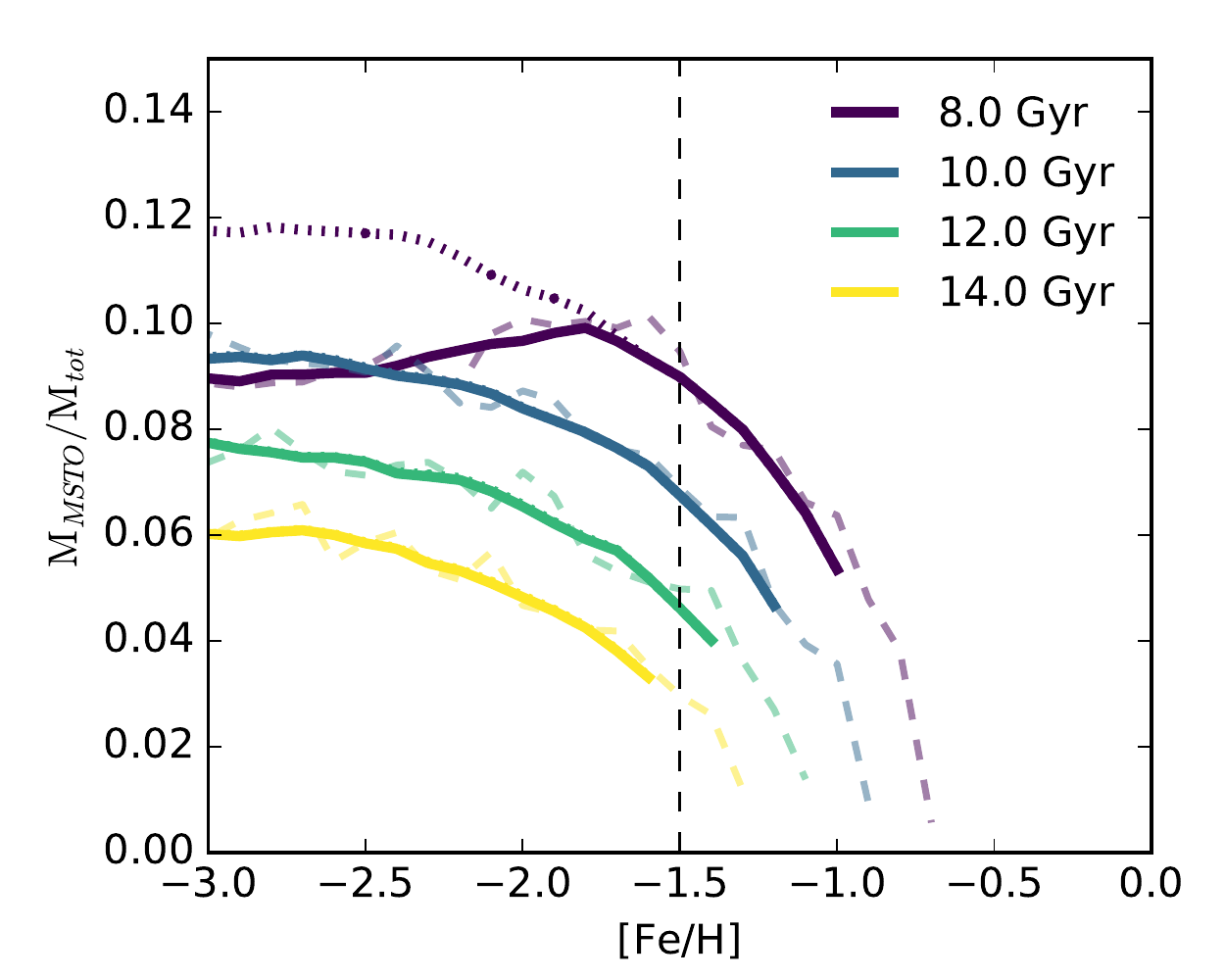}
    \caption{Fractional contribution of the mass from the colour cut of $0.15 < (g_0 - i_0) < 0.4$ to the total mass of the stellar simulated stellar population. The darker lines show the smoothed contours and the lighter dashed lines show the true values in each 0.1 dex metallicity bin. The dotted lines show the mass ratio for a colour cut of $0 < (g_0 - i_0) < 0.4$, and shows the degree of truncation for the MSTO as it extends below $(g_0 - i_0) = 0.15$.}
    \label{frac_MSTO}
\end{figure}


\subsection{Corrections for the halo sample selection}
\label{sec:halo_biases}

For the colour cut used to select the halo sample, there is a metallicity bias that preferentially selects metal-poor stars over metal-rich stars, since the turnoff point of the MSTO is significantly redder for metal-rich populations than for metal-poor ones. This can be clearly seen by the shifting isochrones in the grey region of Figure \ref{CMD_distance}, as a larger fraction of the isochrones fall outside of this region with increasing metallicity. In addition, younger populations will have a bluer turnoff than older ones. In order to investigate this effect and its influence on the MDF, we use PARSEC simulated stellar populations with a total mass of 10~000 M$_\odot$ across a range of metallicities to compute the mass fraction selected by these colour cuts to the total mass of the stellar population. Figure \ref{frac_MSTO} shows the metallicity dependence for the $0.15 < (g_0 - i_0) < 0.4$ colour cut, for a series of ages. The lighter dashed lines show the true values for each bin, and the darker lines are smoothed with a mean kernel and a width of 7 bins. The dotted lines (only visible for the 8 Gyr population) represent a colour cut of $0 < (g_0 - i_0) < 0.4$, and show the result when the entire MSTO is included at all ages. 

For the 8 Gyr simulated population we see the effect of a the tip of the MSTO falling outside of our selected colour range ($g_0-i_0 < 0.15$), shown by the deviation of the solid purple line from the dotted line at $\feh < -1.7$ in Figure \ref{frac_MSTO}. For older ages the MSTO is not truncated and the dotted lines lie underneath the solid lines. Therefore, even though we are not able to reliably assign metallicities to stars bluer than $(g_0-i_0) = 0.15$ with \pris, this will only result in a (small) bias at low metallicities even if there is a significant population of the halo that is younger than 10 Gyr. Furthermore, at all ages the shape of the curves in Figure \ref{frac_MSTO} show that the relative mass selected decreases with increasing metallicity, as this colour selection preferentially selects the (blue) MSTO stars of metal-poor populations. At the most metal-poor end the trend flattens, as the isochrones begin to pile up on each other, i.e. the colour selection is largely unbiased at $\feh < -3$. On the right side of the plot, the metal-rich MSTO populations are increasingly suppressed as they fall outside of the colour selection, i.e. $0.4 < (g_0-i_0)$. Table \ref{corr_frac} in the Appendix gives correction factors for this effect at each metallicity bin. The factors are computed for each age separately by scaling each bin up to the value at $\feh = -3$. The correction we apply to the halo sample is the mean of the 10 and 12 Gyr populations, and is given in the rightmost column of Table \ref{corr_frac}.

 When making the MSTO cuts we also investigate whether using redder broad-band colours to make the MSTO selections could reduce the correction factors, since isochrones of different metallicity do not show as strong of a colour difference as they do with bluer colours. Recomputing Figure \ref{frac_MSTO} for $(i_0-z_0)$, we find that the curves are significantly flatter in the metal-rich regime, indicating a decrease in the metallicity bias when using these colours instead of $(g_0 - i_0)$ colours. However, at the limiting magnitude of $g_0=20$, uncertainties in the $g$ magnitudes are not larger than 0.03, while uncertainties in the $z$ magnitudes in SDSS can be as large as 0.15. Given that in $(i_0-z_0)$ space the MSTO is more compressed, these uncertainties are equal in size to the entire width of the MSTO, and are therefore too large to make a reliable MSTO selection. For this reason, we choose to use the $(g_0 - i_0)$ colours to select the MSTO.



Finally, this colour range also includes horizontal branch (HB) stars (shown in the top part of Figure \ref{CMD_distance}), which are significantly brighter and more distant than the turnoff stars. Using the PARSEC simulated populations, we calculate the mass contribution of HB stars compared to the mass at the rest of the turnoff, and find that in the colour range of $0.15 < g_0 < 0.4$, the relative contribution of HB stars is at most $\sim 0.2\%$ for the 8 Gyr population, and less than this for the older populations. Therefore, we conclude that the HB stars constitute only a very small amount of the sample and subsequently that distance cross contamination between bins is small, resulting in no noticeable effect on the computed MDFs.

\section{Substructure in the halo MDF}\label{sec:mdf}
\subsection{The MDF in different directions}
\label{subsec:direction}

\begin{figure*}
\includegraphics[width=\textwidth]{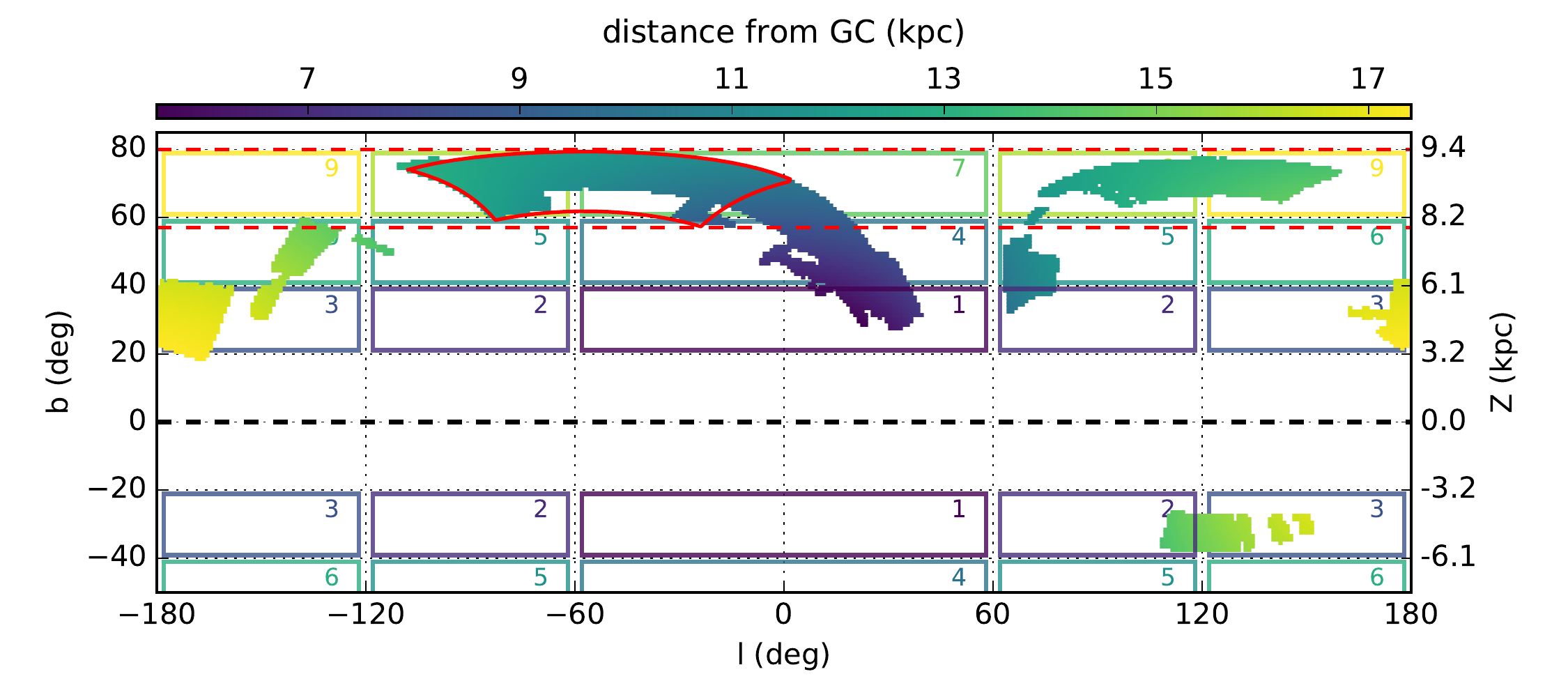}
\caption{The $\sim 2~500\, \mathrm{deg}^2$ \pris footprint, colour coded by the distance to the Galactic center for the halo sample, assuming a heliocentric distance of 9.5 kpc. The coloured squares and the corresponding numbers separate the footprint by latitude and longitude, and each coloured region corresponds to the MDF of the same colour in Figure \ref{mdf_regions}. The region outlined in red shows the region selected to contain the Virgo overdensity, and the red dashed lines are the equal latitude comparison sample.}
\label{distance_gc}
\end{figure*}

We would like to further assess the impact of observational direction on the resulting MDF of the sample. Figure \ref{distance_gc} shows the footprint of the \pris survey on the sky, coloured by distance to the Galactic center. On the left hand y-axis is the Galactic latitude, and on the right hand axis the corresponding distance above or below the Galactic plane (Z = d $\cdot\, sin$(b)). The distances shown are computed to the approximate peak of the halo sample distance distribution of 9.5 kpc, as shown by the black histogram in the left panel of Figure \ref{distance_hist}. To convert from heliocentric distances to Galactocentric, we use a solar position of $(X, Y, Z) = (8.2, 0, 0.011)$ kpc. We separate the sky into 9 regions, made by the intersection of 3 regions in longitude and 3 regions in latitude. The longitude regions are each 60 degrees, and are mirrored across the Galactic center, and the latitude regions are each 20 degrees and are mirrored across the Galactic plane. The 9 regions are shown by the coloured squares and labeled by number in the Figure. Figure \ref{mdf_regions} shows the corresponding MDFs for the halo sample in each of these regions, coloured according to the corresponding region in Figure \ref{distance_gc} and with the number of stars making up the sample also listed in the top left corner. These are plotted over the grey, filled histogram representing the full halo sample for comparison. We see that for most locations in the footprint, the MDF is invariant, meaning that the relative contribution of the disc and halo does not change with varying latitude or longitude, suggesting that the halo cuts were effective in selecting a pure halo sample. Additionally, it shows that the halo is relatively well-mixed in metallicity at these distances. Regions 2, 5, 6 and 9 show some small scale deviations from the total distribution, but these are all regions which have relatively small numbers of stars and thus these features may be a result of insufficient statistical sampling rather than real substructures. In region 3, however, we see a significantly stronger contribution at the metal-rich end, suggesting that at low galactic latitudes there is an additional contribution from a higher metallicity population. 

\subsection{The Monoceros ring}
\label{subsec:monoceros}
Region 3 is at relatively low Galactic latitude and in the direction of the Galactic anticenter. In this observational direction, a substantial amount of substructure has been previously observed, including the Monoceros Ring \citep{Newberg02}, A13 \citep{Sharma10}, TriAnd \citep{Majewski04,Rocha-Pinto04}, the Anti-Center Stream (ACS) and the Easter Banded Structure (EBS) \citep{Grillmair06,Grillmair08,Li12}. Although these structures were originally thought to be disrupted debris from accreted satellite galaxies, some recent studies have shown that some or all of these structures may be the result of kicked-up disc material, given that their chemistry and kinematics are consistent with a disc origin \citep[e.g.][]{Li17, Bergemann18,  Sheffield18}. The peak of the distance distribution of the halo sample coincides with the distance to these structures (d $\sim 10$ kpc), suggesting that these are indeed responsible for the observed metal-rich component in the MDF. Despite being able to differentiate this structure from the rest of the halo based on its metallicity distribution, we cannot from the MDF alone constrain whether this indeed is consistent with kicked-up disc material or debris from an accreted satellite.

To further investigate this metal-rich feature, we separate region 3 into high latitude ($|b| > 30^{\circ}$) and low latitude ($|b| < 30^{\circ}$) stars. The MDFs of each of these latitude ranges are shown in the bottom left panel of Figure \ref{mdf_regions}. From this, we see that for stars with $|b| > 30^{\circ}$, the metal-rich component predominantly disappears, and the MDF is consistent with that of the rest of the footprint regions. We therefore choose to cut out this region from the halo sample and keep only the stars with $|b| > 30^{\circ}$, to minimize potential contamination of metal-rich stars from the disc and its associated structures.

\subsection{The contribution of the disc}
\label{sec:disc_contribution}
\begin{figure}
        \centering
	\includegraphics[width=\columnwidth]{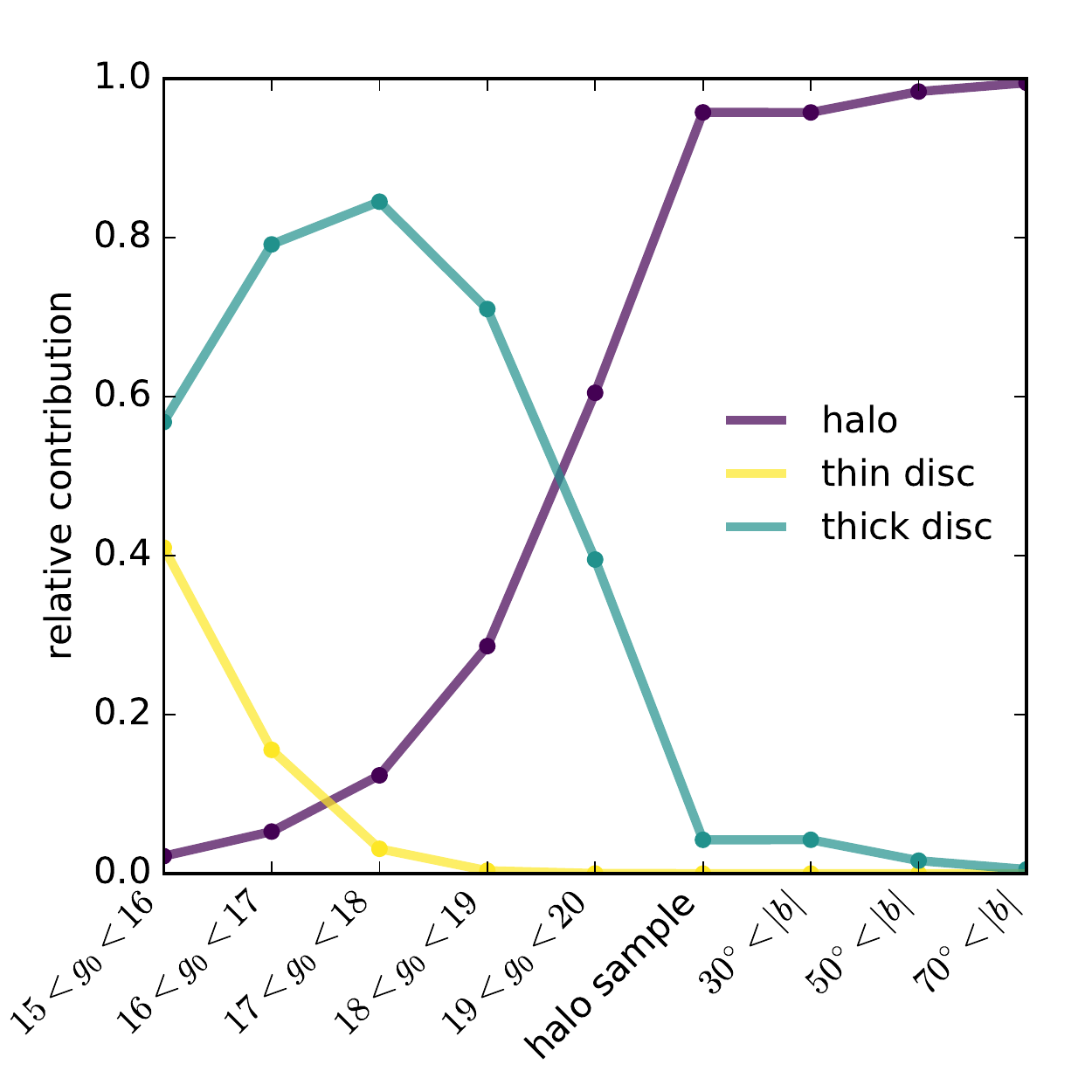}
    \caption{Relative contribution of the thin disc, thick disc and halo for the different distance samples when selected in the Galaxia mock catalogue. The magnitude ranges refer to the MSTO samples, with the colour cut $0.15 < (g_0 - i_0) < 0.6$, and the halo sample refers to the sample made with $0.15 < (g_0 - i_0) < 0.4$ and $19 < g_0 < 20$, as described in the text. The latitude cuts are applied to the halo sample.}
    \label{disc_contribution}
\end{figure}

\begin{figure*}
\includegraphics[width=\textwidth]{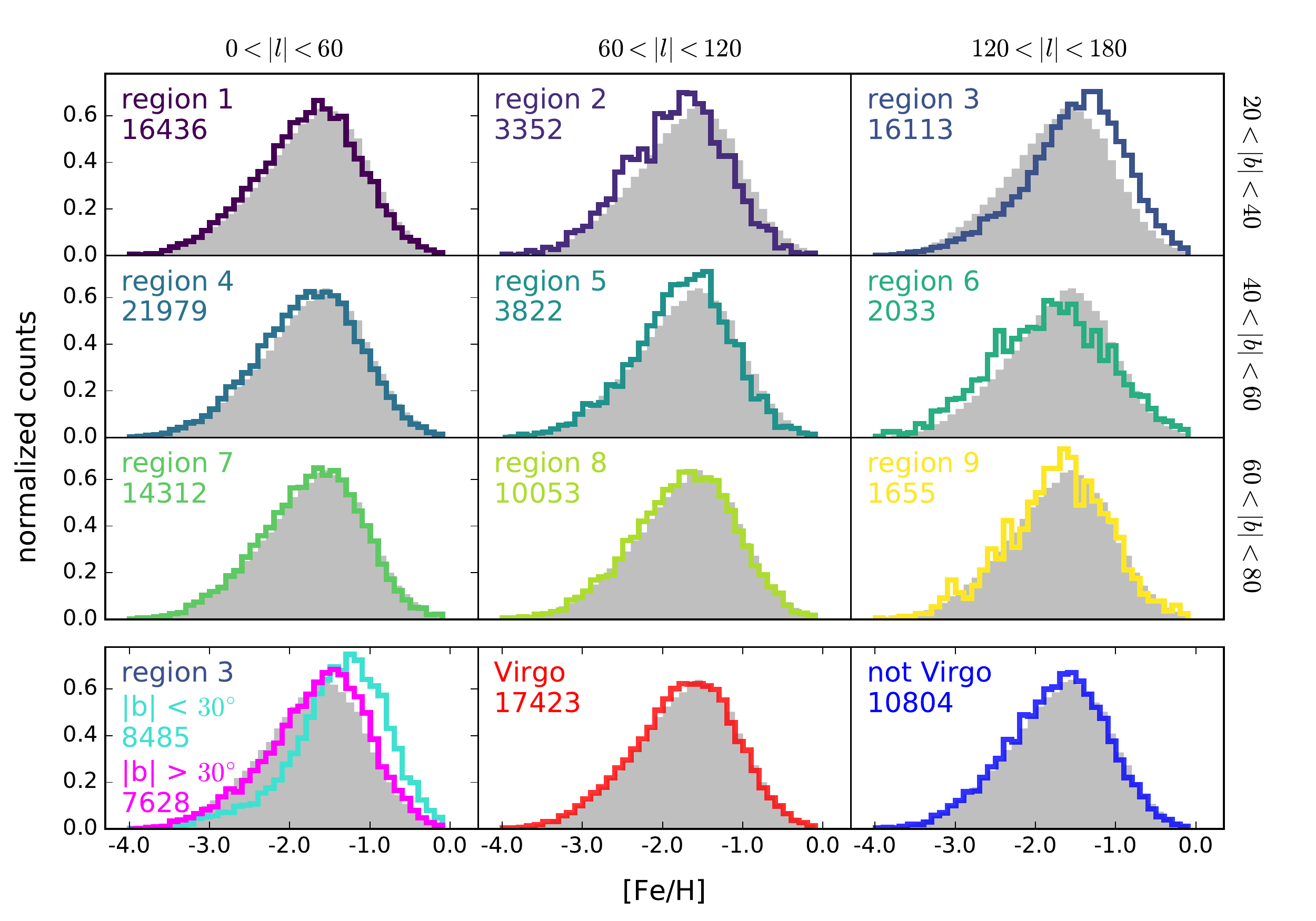}
\caption{MDFs for the 9 different regions described in the text, with colours that correspond to the numbered regions depicted in Figure \ref{distance_gc}. The grey filled histogram is the MDF of the total, GMM corrected halo sample for comparison. The bottom left panel shows the MDFs for region 3, separated into a high and low latitude sample (fuchsia and cyan, respectively), and the other two panels in the bottom row show the region selected to contain Virgo in red (the selected region is outlined in Figure \ref{distance_gc}), and the rest of the stars at the same Galactic latitude as a comparison are shown in blue.}
\label{mdf_regions}
\end{figure*}

We investigate the possible degree of disc contamination in the halo sample further by using the Galaxia mock catalogue described in Section \ref{c4sec:distances}. With this model, we can identify which component of the Galaxy a given star belongs to. Figure \ref{disc_contribution} shows the relative contributions of the thin disc, thick disc, and halo for each of the MSTO selected samples. For the brighter samples with $g_0 < 19$, the sample is dominated by disc stars, and for the $19 < g_0 < 20$ sample there are slightly more halo stars than disc stars. For the halo sample, the contamination from the disc decreases significantly as compared to the faintest MSTO sample. This is due to the greater minimum distance and also due to the removal of metal-rich stars with the narrower colour cuts when selecting the MSTO. Adding latitude cuts further increases the purity of the halo selection until at a latitude of $70^{\circ}<|b|$ the sample consists of almost entirely halo stars. Although this model represents a simplified version of the Galaxy, with an axisymmetric disc devoid of perturbations and substructures from interactions with accreted satellites, this still offers insight into the relative contributions of the selections to the purity of the sample. Therefore, due to the distance range probed by the selected halo sample, the contribution of foreground metal-rich disc stars is likely to be low, although there may still be some component of high latitude, kicked-up disc material, as discussed in Section \ref{subsec:monoceros}.

\subsection{The Virgo overdensity}
\label{subsec:virgo}
In addition to investigating the effect of direction in Figures \ref{distance_gc} and \ref{mdf_regions}, we also take a closer look at the largest Galactic substructure in our footprint that is probed by our distance range: the Virgo overdensity. Although the specific nature and origin of this structure is still not well understood, it is known that the Virgo overdensity is located at a distance of 8 - 25 kpc \citep[e.g.,][]{Lokhorst16}, and that it is likely comprised of overdense regions of halo stars. Located at approximately RA = 190$^{\circ}$ and Dec = -5$^{\circ}$ and spanning a large region of sky of over 1~000 deg$^2$, a significant region of the overdensity is located within the \pris footprint. The current data set is therefore well suited to exploring the MDF of this structure. Although the Sagittarius stream is also prominent over our footprint, its vast leading and trailing arms are too distant to be sufficiently seen in the sample presented in this work \citep[see for example][]{Law+Majewski10}. We therefore select a region around the Virgo overdensity, and compare its metallicity distribution to another region of the \pris footprint at the same Galactic latitude that does not include Virgo. The region selected to contain Virgo is 180$^{\circ}$ < RA < 210$^{\circ}$ and -1$^{\circ}$ < Dec < 20$^{\circ}$, and is demarcated in red in Figure \ref{distance_gc}, and the comparison sample is shown by the red dashed lines. The MDFs of these regions are shown in the bottom middle and bottom right panels of Figure \ref{mdf_regions}. Although there are minor fluctuations from the underlying halo sample, these are not significant enough to draw any conclusions. Therefore, this indicates that the Virgo overdensity metallicity is not very different from that of the general stellar halo at these Galactic latitudes.


\section{Discussion}
\label{sec:disc}

So far, we have produced an MDF for a carefully selected halo sample, with $0.15 < (g_0 - i_0) < 0.4$, $19 < g_0 < 20$, $|b| > 30^{\circ}$ and a cut in $(u_0 - g_0)$ vs $(g_0 - i_0)$ space to remove young disc stars. We calculated with isochrones and a Galaxia model that this sample spans $\sim 6-20$ kpc, a distance range which nicely probes the inner Galactic halo. This sample is not free of selection biases, as we discussed in Section \ref{sec:halo_biases}, but these are mostly present at the metal-rich end, and constitute moderate corrections that we have applied at $\feh < -1.5$. 

The slope of the metal-poor regime, particularly at $\feh < -1.5$, is important in that it constrains the relative ratios of the most metal-poor stars in the Galaxy, which in turn encode the rate of chemical enrichment at early times. This is useful for building chemical enrichment models and for constraining the physical processes driving them. This also helps to better understand the role of metal-poor stellar populations in the formation of the Galactic halo, such as accreted dwarf galaxies and globular clusters.

\subsection{Biases in the photometric metallicities}
\label{sec:phot_bias}

The \pris survey is fairly unbiased in the stars that it observes. However, there are still various inter-dependencies on the colour, magnitude, and derived metallicities. We list the most important effects below, and discuss their possible effects on the MDF in the following section.

\begin{figure}
        \centering
	\includegraphics[width=\columnwidth]{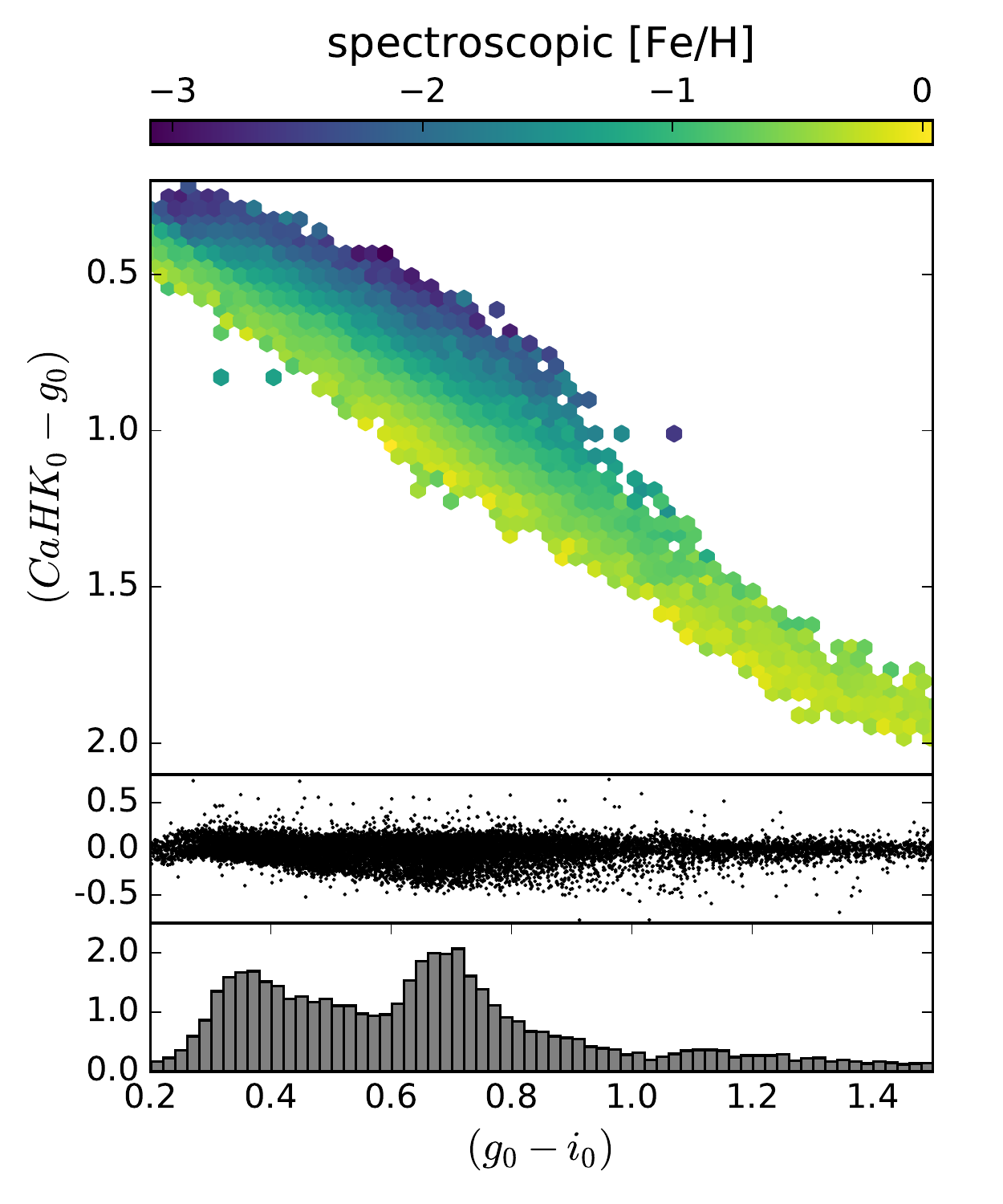}
    \caption{Top panel: Colour-colour plot of the SEGUE spectroscopic sample, colour-coded by SEGUE metallicities. Middle panel: $(CaHK_0 - g_0)$ colour subtracted by the median value per $(g_0 - i_0)$ colour bin. Bottom panel: histogram of the distribution of the top panel.}
    \label{CaHK_g_vs_g_i}
\end{figure}

\begin{itemize}
    \item {\textit{Dependence of the metallicity sensitivity on colour:} Figure \ref{CaHK_g_vs_g_i} shows the overlapping sample of SDSS/SEGUE stars with the \textit{Pristine} sample (using the same quality cuts on the \textit{Pristine} photometry and the SDSS/SEGUE spectra as described in Section 3 of \citealt{Starkenburg17}). The colour-coding by metallicity nicely illustrates the metallicity sensitivity of the survey. This figure also shows that at bluer colours the range of metallicities span a narrower range in $(CaHK_0 - g_0)$ than at redder colors, meaning that for the same uncertainties in photometry there will be a larger uncertainty in metallicity, and consequently a larger contamination across metallicity bins. In addition, the histogram in the bottom panel of Figure \ref{CaHK_g_vs_g_i} shows that most of the stars are concentrated at $(g_0 - i_0)$ < 1.0. The population redder than that is mostly dominated by metal-rich disc dwarfs.}
    \label{list_1}
    \item {\textit{Dependence on the depth of the survey on metallicity:} Due to the nature of the narrow-band \textit{Pristine} CaHK filter, stars that are metal-poor will have smaller absorption lines and thus appear brighter in the photometry. This results in a slight difference in brightness for the most metal-poor stars as compared to the metal-rich ones. The middle panel of Figure \ref{CaHK_g_vs_g_i} shows that on average, the more metal-poor stars will be $0.2-0.5$ mag brighter in the \pris narrow-band filter than metal-rich stars. This will lead to the metal-poor stars probing a slightly larger volume. However, imposing the colour cut of $0.15 < (g_0 - i_0) < 0.4$ used to select the halo sample reduces this metallicity dependant magnitude discrepancy to $0.1-0.2$ mag, which is on the order of the uncertainty inherent in determining these distances with isochrones.} 
    \label{list_2} 
    \item{\textit{Dependence of measurement uncertainties on magnitude:} Another effect results from the uncertainties on the photometric flux being larger at fainter magnitudes. At the bright end, the uncertainties in both the CaHK and the SDSS broad-band magnitudes are small, but at the faint end $\sigma_{CaHK} >> \sigma_{SDSS}$, such that it is largely the CaHK uncertainty that drives the overall metallicity uncertainty. The photometric uncertainties as a function of magnitude for the sample are summarized in Table \ref{c4:unc_table}. The larger uncertainties at the faint end are partially mitigated with the GMM rescaling, especially since we construct the GMM based on a fainter sample, resulting in a model that takes into account the larger uncertainties in the spectroscopic data set.} 
    \label{list_3} 
    \item{\textit{Dependence of contamination fraction as a function of metallicity:} As previously mentioned in the introduction, there are far more metal-rich stars than metal-poor stars in the halo MDF, meaning that more metal-rich stars will scatter into the metal-poor regime than the other way around, resulting in a relatively higher contamination fraction at low metallicities. This is largely corrected for by the GMM rescaling, but only in so far as metal-rich stars that scatter into the metal-poor end can be identified by medium resolution spectroscopy.}
    \label{list_4}
    \end{itemize}


\begin{table}
\centering
\caption{Uncertainties as a function of magnitude for the \pris CaHK and SDSS broad-band magnitudes.}
\begin{tabular}{cccc}
\hline
\textbf{}          & \multicolumn{3}{c}{median uncertainties}         \\ \hline
Magnitude & g & i & CaHK \\ \hline
g = 15             & 0.003            & 0.004            & 0.004               \\ 
g = 16             & 0.004            & 0.004            & 0.006               \\ 
g = 17             & 0.005            & 0.005            & 0.01                \\ 
g = 18             & 0.006            & 0.006            & 0.02                \\ 
g = 19             & 0.01             & 0.01             & 0.05                \\ 
g = 20             & 0.02             & 0.02             & 0.1                 \\ 
\hline
\label{c4:unc_table}
\end{tabular}
\end{table}

\subsection{Previous determinations of the halo MDF}


The MDF of the halo produced from \pris data is shown in Figure \ref{mdf_lit_comp_log}. The grey histogram is the raw \pris MDF, the cyan is after rescalling with the GMM and the black histrogram is corrected with both the GMM and the correction factors for the colour cut from Table \ref{corr_frac}. Application of the GMM correction to the halo sample has two main effects on the MDF, namely that it shifts the peak of the distribution to slightly higher metallicities, and removes contaminants from the metal-poor tail and shifts them to higher metallicities. The corrections for the colour cut do not significantly change the slope of the metal-poor tail of the MDF ($\feh < -2$), but modestly enhance the relative numbers of stars in the bins at higher metallicities.

The right panel of Figure \ref{mdf_lit_comp_log} shows a comparison of several MDFs from the literature that are presented in this paper, and the left panel shows the same but with a log scale to better emphasize the differences in the slopes of the metal-poor regime. Each distribution is scaled to the \pris distribution at $\feh = -2$, to allow for an easier comparison of the slopes. The purple, yellow and green histograms are the MDFs presented in \citet{An15}, \citet{Allende14} and \citet{Schoerck09}, respectively, and the blue dashed line is the MDF expected from the simple chemical enrichment model presented in \citet{Hartwick76}, assuming an effective yield of $y_{\mathrm{eff}}$ = -1.6. The MDF of \citet{Ryan91} is not shown in this Figure, but is in good agreement with the \citet{Hartwick76} model. In this section, we first describe each MDF and how it was produced, and then discuss how it compares to our
results.

\subsubsection{The chemical evolution model of \citet{Hartwick76}}
Using a sample of 60 halo globular clusters, \citet{Hartwick76} modifies a simple chemical evolution model \citep{Searle+Sargent72, Pagel+Patchett75} to describe the MDF. Starting from a closed-box of primordial gas, this model is based on a simple parametrization of the rate at which the ISM is enriched in heavy elements by exploding stars. The model assumes instantaneous recycling and a constant IMF, and can be described using a single free parameter known as the effective yield, $y_{\mathrm{eff}}$. This parameter represents the ratio of the mass of heavy elements ejected, to the mass locked up in long-lived stars or remnants, and also sets the location of the peak of the distribution. Following \citet{Ryan91} we choose $y_{\mathrm{eff}}$ = -1.6. Under the assumption that the halo globular cluster population traces the field star population, we can use this model to approximate the idealized MDF of the Milky Way halo. 

\subsubsection{SDSS sample \citep{Allende14}}
\label{sec:AP14}
\begin{figure*}
	\includegraphics[width=\textwidth]{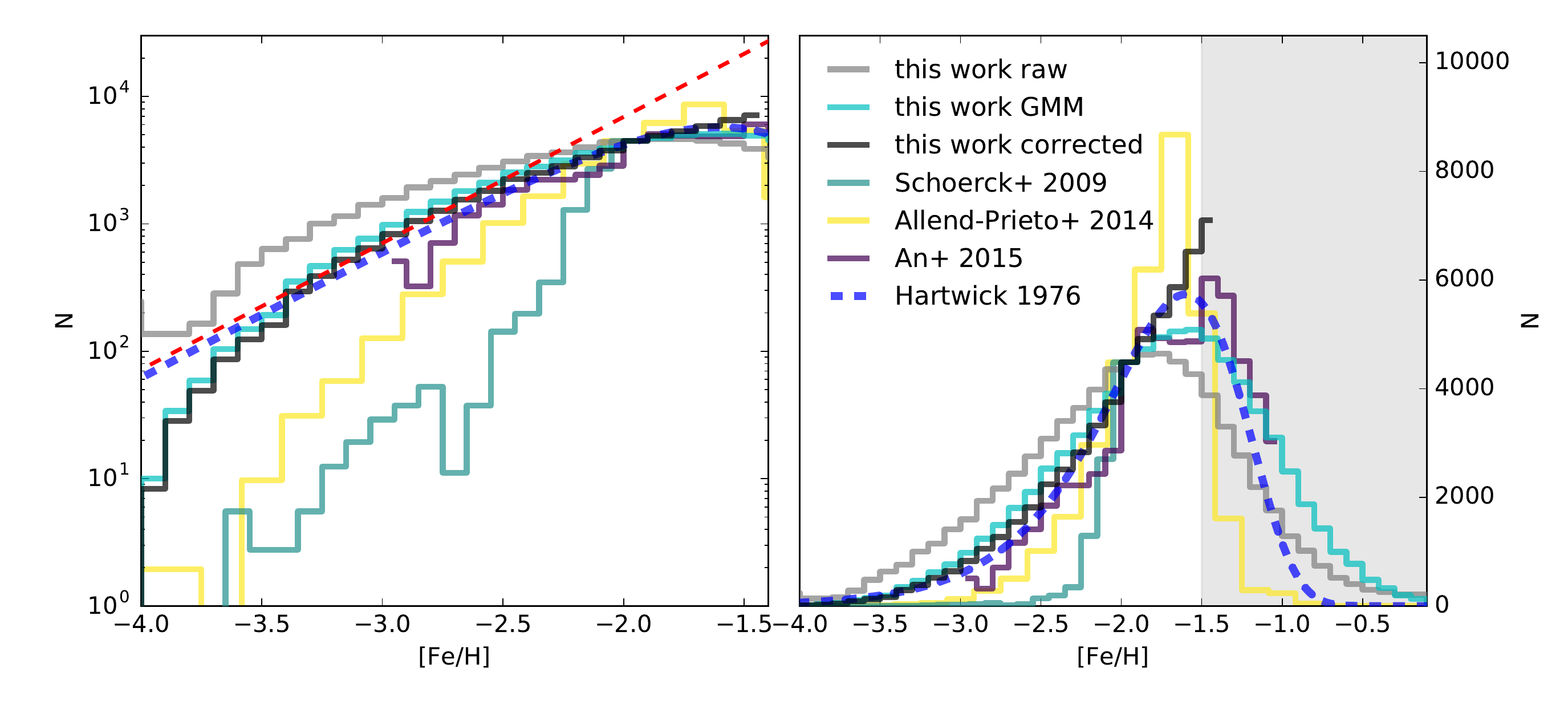}
    \caption{Comparison of the \emph{Pristine} MDF to previous works. The right panel shows the MDFs over the full metallicity range to illustrate the global shapes and location of the peaks, while the left panel shows only the metal-poor end and is plotted with a log scale to emphasize the differences in the metal-poor slopes. The grey shaded region shows where the corrections to the \pris MDF become large and uncertain, at $\feh > -1.5$. The raw MDF of the \pris halo sample is shown in grey, the GMM corrected in cyan, and the GMM/colour corrected in black. The MDFs from \citet{An15}, \citet{Allende14} and \citet{Schoerck09} are shown in purple, yellow, and green respectively, and the \citet{Hartwick76} model is shown as the blue dashed line. The red dashed line is an exponential fit to the \pris corrected sample in the metallicity range of $-3.4 < \feh < -2.5$.}
    \label{mdf_lit_comp_log}
\end{figure*}

\citet{Allende14} use a sample of spectroscopically selected F-type stars from the SDSS survey (in particular those targeted in the BOSS survey as flux calibrators) to determine a metallicity distribution of the Milky Way halo. The MDF shown by the yellow histogram in Figure \ref{mdf_lit_comp_log} is produced using a sample of $\sim 16~000$ F-type MSTO stars selected with $2.5 < \logg < 4.4$ and $g_0 > 17$. Most of the observed stars are at high Galactic latitudes, and they estimate them to probe a distance range from 5 kpc out to a few tens of kpc. They note that this sample is biased due to colour cuts used to select the BOSS spectrophotometric stars (summarized in their Table 1). To correct for these, they compute photometric colours for a grid of spectral energy distribution models covering a range of stellar parameters and identify which ones make it through the colour cut and are in the temperature range of 5~600 < \teff\, < 6~500 K. They then rescale the MDF to account for the missing populations, making the simplifying assumption that all stars are evenly distributed over the stellar parameter space. Although this slightly boosts the number of stars at the peak of the distribution at $\feh = -1.6$, it has very little effect on the tails of the distribution. The corrections are expected to be most significant for higher metallicities, as at these metallicities few stars with typical turnoff temperatures and gravities are selected in BOSS \citep[as is also apparent from the results in Table 1 of][]{Allende14}.

\subsubsection{HES sample \citep{Schoerck09}}
\citet{Schoerck09} estimate the slope of the MDF at the metal-poor end by correcting the observed MDF from the HES survey for selection effects. For the correction, they create an MDF following \citet{Hartwick76}, replace all the $\feh$ values with pairs in their observables KP \citep[an index on the Ca K absorption line following][]{Beers99} and color $(B-V)_0$, convolve them with Gaussian uncertainties, and assess the completeness in $\feh$ vs $(B-V)_0$ space after applying the HES selection cuts. The result of this exercise is that they find HES to be essentially complete below $\feh= -3.5$, but the corrections become quite significant at higher metallicities. For example, at $\feh=-2.05$ they apply corrections of a factor of $10-80$ (depending on colour). They build the MDF based on a sample of 1638 HES stars with available spectroscopic follow-up, and scale it up to include all 3~439 stars from the total candidate sample.


\subsubsection{The sample of \citet{Ryan91}}
\citet{Ryan91} present the metallicity distribution of the local halo from 372 subdwarfs selected in proper motion studies with space velocities relative to the LSR > 250 km s$^{-1}$. They find a very good fit to the \citet{Hartwick76} model using $y_{\mathrm{eff}} = -1.6$. The excellent fit does not include the 10\% most metal-rich stars in the sample, but they speculate that this might be contamination from the disc. The model would predict 13 stars with $\feh <-3$, whereas 6 are observed, this difference is however too small to be significant for their sample size.

\subsubsection{The sample of \citet{An15}}
\label{subsubsec:An15}
\citet{An15} derive a halo MDF photometrically from main-sequence turnoff stars observed in the extra deep SDSS stripe 82, and estimate that their sample covers a distance range of $5-10$ kpc. They find excellent agreement with the work of \citet{Ryan91} mentioned above. In addition to a single \citet{Hartwick76} or Gaussian model fit, they also provide a double Gaussian fit to their derived MDF, finding a peak at $\feh =-1.4$ and a second one at $\feh =-1.9$. The extra Gaussian component is motivated by a difference in kinematics for the stars in both populations, where the second population shows a larger number of retrograde orbiting stars.

\subsection{Comparison to the literature}
\label{subsec:lit_comparison}
\subsubsection{The number of metal-poor stars}
\label{subsubsec:number_metal_poor}
Figure \ref{mdf_lit_comp_log} shows that there are distinct differences between the various MDFs in the relative numbers of stars they contain at $\feh < -2$. Most notably, the \pris halo sample shows the largest number of stars, even more than the \citet{Hartwick76} model, in the range $-3.4 < \feh < -2.2$. The photometric sample of \citet{An15} shows a pronounced bimodality, with the metal-poor component showing a similar abundance of stars to \pris for $\feh < -2.3$, but it begins to drop off at $\feh < -2.6$, presumably as the metallicity calibration becomes less reliable. Both of these samples are in reasonably good agreement with the model. The SDSS/BOSS spectroscopic sample of \citet{Allende14} shows a smaller contribution with respect to the photometric samples at $\feh < -2$, and the HES sample of \citet{Schoerck09} shows the lowest relative contribution of metal-poor stars of all of the shown MDFs.

From the left panel of Figure \ref{mdf_lit_comp_log}, we can see that the shape of the \pris halo sample MDF is not linear in log space, and therefore is not well described by a single exponential curve. Instead, the observed distribution is much more suited to a fit with 3 separate exponentials, in the metallicity ranges of $\feh < -3.5$, $-3.4 < \feh < -2.5$, and $-2.5 < \feh < -1.5$. The best fit line in the metallicity range of $-3.4 < \feh < -2.5$ is shown as the red dashed line in the left panel of Figure \ref{mdf_lit_comp_log}, and has a slope of $\Delta$(LogN)/$\Delta\feh = 1.0 \pm 0.1$. Below this metallicity range the slope becomes substantially steeper with a value of 2.0 $\pm$ 0.2. 

Using a spectroscopic sample of metal-poor halo giants selected from the \emph{SkyMapper} survey, \citet{DaCosta19} derive a slope of the MDF of $\Delta$(LogN)/$\Delta\feh = 1.5 \pm 0.1$ for a metallicity range of $-2.75 < \feh < -4$. The MDF they derive is well fit by a single exponential for the entire metallicity range, but drops off steeply at $\feh = -4$. Fitting a single exponential in the same way naively to the \pris MDF also yields a slope of 1.5 $\pm$ 0.1, but as discussed above, this does a poor job of describing the distribution. Therefore, the slope in the metal-poor regime from the \pris halo sample is somewhat flatter, consists of more components and has a drop off at a higher metallicity ($\feh < -3.5$) than the MDF described in \citet{DaCosta19}. The slopes of both of the spectroscopic samples are also steeper than the slope of the \pris halo sample for $\feh < -2$, particularly for the \citet{Schoerck09} sample which shows the steepest slope in the metallicity range of $-2.5 < \feh < -2$ of all the MDFs.

Interestingly, the slope of the \pris MDF between $-3.4 < \feh < -2.5$ matches very well to the slope of the \citet{Hartwick76} simple chemical evolution model at $\feh < -2$. This breaks down at $\feh < -3.5$, however, where the \pris MDF drops steeply. This also happens to be the metallicity below which the \pris photometric metallicity determination becomes less reliable, and where there are very few stars in the spectroscopic sample with which to fit the GMM correction. Nevertheless, this steep decrease is seen in the other MDFs as well, which all show a drop between $-4 < \feh < -3.5$, whereas the model does not. At these very low metallicities -- and at the earliest epoch of star formation -- chemical enrichment is dominated by a small number of supernovae explosions. Therefore, star formation is likely to proceed in a very inhomogeneous and stochastic way, deviating more strongly from the assumptions of the model that the ISM is well-mixed and homogeneous. Additionally, the model does not account for channels of enrichment from external sources, which could very quickly enrich the interstellar medium and result in a steeper slope at the lowest metallicities. In \citet{Schoerck09}, they note that despite the discrepancy between the slope of their metal-poor MDF and the \citet{Hartwick76} model, their MDF better matches theoretical models which include a critical metallicity for low-mass star formation at $Z_{cr} = 10^{-3.4}Z_{\odot}$ \citep{Salvadori07}. It should be noted, however, that \citet{Schoerck09} rely on rescaling the models to match their selection function. Nevertheless, this motivates the exploration of further modifications of these simplified models by incorporating more complex physics (e.g. gas accretion, outflow and inhomogeneous mixing) which may improve the agreement between the models and the data.

 
\subsubsection{Selection effects and metallicity bias}
\label{subsubsec:phot_vs_spec}
Each of the MDFs discussed above are selected in different ways and from different samples. \citet{Ryan91} and \citet{Hartwick76} describe the local halo, whereas \citet{An15}, \citet{Schoerck09}, \citet{Allende14} and the current work all target the "distant" halo, by selecting stars at larger distances from the Galactic plane, albeit each exploring a slightly different distance range and thus different Galactic environments. More importantly than just the differences of environments probed, is that each of these samples come with their own selection functions and biases. Despite all of these works applying careful corrections to the MDFs, they still do not agree fully with one another, particularly at the metal-poor end. 

As previously discussed, the greatest difficulty in producing an MDF is to correct for metallicity dependant biases and selection effects incurred from the selection of a halo sample. Spectroscopic samples are typically more affected by completeness issues, since stars are specifically selected for follow-up observations, whereas photometric samples observe stars more ubiquitously. In addition, spectroscopic samples may be designed to target metal-poor stars, and thus have an implicit bias towards stars of low metallicity which must be corrected (such as is the case for HES for example). This correction process is highly non-trivial, and must be based on some models or assumed knowledge of the underlying stellar populations, and an over correction could lead to an underestimate of the number of metal-poor stars. In contrast, a photometric approach is likely much more complete (in so far as it is complete within the probed magnitude range), but is more susceptible to contamination due to the larger uncertainties associated with photometric metallicity determinations. These uncertainties may cause a broadening of the distribution, and an increase in the number of stars in the wings (i.e. at the metal-poor and metal-rich ends). We have made an effort to correct for this contamination by applying a statistical rescaling of the MDF with a GMM fit to a spectroscopic sample, but this is only effective in correcting for erroneous photometric metallicity determinations that are caught by the spectroscopy. Stars that have peculiar features in their spectra which affect their metallicities but that cannot be detected with low/medium resolution spectroscopy (e.g. certain binaries, blue stragglers) will go undetected and thus uncorrected by this approach, and result in an over estimation of the number of metal-poor stars. We have discussed the potential influence of these contaminants in Section \ref{sec:phot_bias}, and although their contribution is likely to be small, this remains a caveat of the current work. 
 
Nevertheless, a photometric sample has other advantages, such as larger sample sizes. Despite the many cuts imposed to make the \pris halo sample, it still consists of $\sim$ 80~000 stars. The next largest sample from the ones discussed above is from \citet{Allende14}, and consists of $\sim$ 16~000 stars, although the $\logg$ cuts imposed for that sample are more comparable to the faintest MSTO sample presented in this work, which consists of $\sim$ 230~000 stars. The narrower $\logg$ sample from \citet{Allende14} that is comparable to our halo sample has < 1~000 stars. This naturally allows for an MDF that is statistically more robust than those made with the somewhat smaller spectroscopic samples.

Ultimately, none of the MDFs presented here are completely devoid of bias or selection effects. It is only through multiple, repeated efforts using a diversity of independent samples and methods, each with their own selection effects and biases, that we may hope to converge on a true determination of the halo MDF. This is particularly promising as we continue to collect large amounts of spectra with large scale surveys, and continue to improve the quality of photometric metallicity samples and distance determinations.
 

\subsection{Comparison with Gaia-Enceladus}
\label{subsec:implications}

The MDF presented for the halo sample, with its peak $\feh = -1.6$ is qualitatively consistent with the MDF presented for the Gaia Enceladus structure in \citet{Helmi18} (histogram in their Figure 2), although the MDF presented in this work has a much stronger tail of metal-poor stars then the one presented for Gaia Enceladus. It should be noted that the MDF presented in \citet{Helmi18} is produced using a sample of APOGEE stars, and no attempt is made to correct for selection effects. Another study conducted by \citet{Gallart19}, select halo stars by taking all stars with large tangential velocities (> 200 km/s) from a local sample ($\lessapprox$ 2 kpc). They further separate these stars based on their metallicity distributions into a blue sequence that is consistent with the Gaia Enceladus population, and a red sequence that they suggest is representative of an "in situ" halo population. They determine the metallicity and age distributions for these population using a linear combination of synthetic isochrones, and find that they are both consistent with very old stellar populations. The photometric MDFs for these two sequences match reasonably well with spectroscopic MDFs produced by cross-matching with the LAMOST and GALAH spectroscopic surveys. The MDF of the blue sequence alone has a peak of $\feh \sim -1.4$ and loosely follows the distribution provided in \citet{Helmi18}, but the MDF of the red sequence is much more metal-rich with a peak at $\feh \sim -0.7$. If we consider these two populations together to make up the halo, then there is a much stronger metal-rich population than is shown in the MDF of the halo sample that we present in this work. However, this may simply indicate differences that result from selection of halo samples in both works as well as the limitations at high metallicity in this work. If the inner halo is indeed dominated by the stars from a single massive accretion event, then the MDF presented here with the distance range of our halo sample should consist mostly of those stars. However, despite qualitative similarities between the distributions, we cannot with the current data corroborate or reject this hypothesis.




\subsection{The duality of the stellar halo}
\label{subsec:duality}
It remains an open question in the field of Galactic archaeology of whether the Galactic halo is composed of a single or multiple components. As discussed in Section \ref{subsubsec:An15}, \citet{An15} derive a halo MDF that is consistent with having two distinct components, and suggest that this supports the dual halo hypothesis. The purple histogram in Figure \ref{mdf_lit_comp_log} shows a fairly clear bimodality, to which \citet{An15} fit two Gaussians with peaks at $\feh = -1.4$ and $-1.9$. They go on to estimate that the local population of halo stars is made up of $35-55\%$ outer halo stars. The \pris MDF does not show such a clear bimodality, but is also clearly not well fit by a single exponential in the metal-poor regime, which may indicate some more complex underlying populations. Similarly, neither of the spectroscopic MDFs from \citet{Schoerck09} or \citet{Allende14} display an obvious bimodality, and neither of those authors discuss a potential bimodality in the MDFs as evidence for a dual halo. Interestingly, however, the halo MDF produced in \citet{Zuo17} with SDSS photometry also does not show a clear bimodality, but they nevertheless decompose it into 3 underlying Gaussian components, one quite metal-rich and consistent with a disc population, and two that they identify as the inner and outer halo. They show this for two samples, one with $-8 < Z < -4$ and $10 < R < 14$ that shows peaks of $\feh = -1.43$ and $-1.92$, and a more distant sample with $-12 < Z < -8$ and $12 < R < 16$ for which they derive peaks of $\feh = -1.51$ and $-2.25$. With our current data, we do not have precise enough distances, or a clean enough selection of halo giant samples to go further out into the halo and see if a well defined break radius between these two halo populations can be identified, although this should be the topic of future work. Therefore, we do not draw a firm conclusion as to whether the MDF presented in this work is consistent with a multi or single component halo.


\subsection{Implications for the formation of globular clusters}
\label{sec:globular_clusters}

In the Milky Way there is an apparent paucity of globular clusters (GCs) at low metallicities. More specifically, of the $\sim$ 160 GCs observed to date, there are none with $\feh < -2.5$ \citep{Harris10}, which is seemingly at odds with the old ages derived for many GCs. In order to gain insight into this discrepancy, we can compare the metal-poor MDF of Milky Way GCs to the MDF of halo field stars to differentiate whether this lack of observed GCs could be the result of a real physical process or simply due to statistical undersampling. From the catalogue of \citet{Harris10}, there are 56 GCs with $\feh \leq -1.5$, 13 with $\feh \leq -2$, and 0 with $\feh \leq -2.5$. Using the relative number of stars at these metallicities from the scaled MDF of this work, we calculate an expected number of GCs to be 56, 26, 10, and 2.8 for $\feh \leq -1.5$, $-2$, $-2.5$ and $-3$, respectively. Taking only the GCs in the Galactocentric distance range of the selected halo sample ($d_{GC} \sim 5-20$ kpc), the counts from the \citet{Harris10} catalogue are 30, 10, 0, and 0, and from \pris 30, 14, 5.4, and 1.5 for $\feh \leq$ $-1.5$, $-2$, $-2.5$, $-3$, respectively. Taking a Poisson distribution with a mean of 5.4, the expected probability of finding no clusters at $\feh < -2.5$ would be 0.5\%, while a Poisson distribution with a mean of 1.5 gives a probability of 22\% of finding no clusters at $\feh < -3$. Therefore, we see that if the GC population were to follow the same MDF as the halo field stars, there is a significant discrepancy between the observed and expected number of GCs at $\feh < -2.5$. This is suggestive that the perceived lack of metal-poor GCs is due to a real truncation of the GC MDF, or alternatively that there are underlying differences in formation mechanisms of GCs with respect to the field stars. A more robust statistical analysis to put these results on a firmer footing, as well as a more in-depth discussion will be presented in Gieles et al., in prep.

Some recent observational efforts have identified low metallicity objects in the faint, low mass regime where globular clusters and dwarf galaxies are difficult to differentiate. For example, the Sylgr stream \citep{Ibata19} contains at least two stars with $\feh = -2.92 \pm 0.06$ \citep{Roederer19}, although it cannot yet be determined whether the progenitor of this stream was a dwarf galaxy or a GC. Another object, Draco II \citep{Laevens15}, has a low metallicity ($\feh = -2.7 \pm 0.1$), and shows both dwarf galaxy and GC like characteristics. In an in depth study of this system, \citet{Longeard18} favour the scenario that Draco II is a disrupting dwarf galaxy, although they leave open the possibility that it could be a GC observed at the very end of its disruption process. Therefore, it remains to be seen whether the perceived lack of low metallicity GCs will be solved by detailed observations in the near future, or if this needs to be explained by theories of GC formation.

\subsection{Future work}

The halo MDF we present here relies on a spectroscopic sample to rescale the photometric metallicities. The \pris spectroscopic follow-up sample, totalling $\sim$ 1~000 stars, was obtained over a three year period using a total of $\sim 165$ nights of observing time on single-slit spectrographs mounted on $2-4$m class telescopes. Although this approach was successful at obtaining a sample with which to explore the data set, it is unfeasable to follow-up the full number of \pris candidates, particularly at fainter magnitudes. The William Herschel Telescope Enhanced Area Velocity Explorer \citep[WEAVE]{Dalton18} is a multi-object spectrograph (MOS), and starting this year, will begin collecting a large number of spectra to survey the northern sky. In a Memorandum of Understanding between the surveys, it is agreed that WEAVE will allocate $\sim$ 6 fibres per deg$^2$ to follow-up \pris metal-poor star candidates as part of its Galactic archaeology program. As the \pris survey footprint currently spans over $\sim$ 5~000 deg$^2$ and is still growing, we estimate that the WEAVE survey will observe a total of $\sim$ 30~000 \pris candidates during its operation. Apart from providing an excellent sample of metal-poor stars in the halo with which to conduct a multitude of Galactic archaeology studies, this will also provide a much larger spectroscopic sample with which to extend the current work. This will allow for an even more detailed investigation of the MDF in various galactic environments as we will have much more information on the dependence of contamination in the samples as a function of colour and magnitude.

Other upcoming MOS facilities and large spectroscopic surveys will continue to provide large numbers of spectra of halo stars that can be used to determine MDFs when their selection functions are not too complex, or with which to correct photometric surveys as in this work. In combination with the next data releases from the Gaia mission including more precise and accurate parallaxes, they will provide an excellent opportunity to study the metallicity distribution function extending further out into the halo and in narrower distance ranges.

\section{Conclusions}\label{sec:conc}

In this paper, we built an MDF of the Galactic halo using a carefully selected halo sample from the \pris survey. The \pris survey currently covers an area of $\sim$ 2~500 deg$^2$ and provides photometric metallicities to $\sim$ 4 million FGK type stars. From this sample, we applied colour and magnitude cuts to select MSTO samples spanning various heliocentric distance ranges between $0.6 - 25$ kpc. We found that there is a gradient showing decreasing metallicity with increasing distance, with a clear transition from a strong disc component at $\feh \sim -0.7$ in the brighter samples to a metal-poor peak at $\feh \sim -1.6$ where the halo dominates in the fainter samples. We then selected a halo sample consisting of $\sim$ 80~000 stars, selected with $0.15 < (g_0 - i_0) < 0.4$, $19 < g_0 < 20$, $|b| > 30^{\circ}$, and a cut in $(u_0 - g_0)$ vs $(g_0 - i_0)$ space to remove young foreground disc stars. This produced a sample covering a heliocentric distance range of $\sim 6 - 20$ kpc, which we confirmed to have a Galactocentric radius of $5 < R_{GC} < 20$ kpc and height above the disc $|Z| > 4$ kpc using a mock \pris sample generated with the Galaxia code. We applied a statistical rescaling based on a sample of \pris follow-up spectroscopy and SDSS/SEGUE stars, to account for contamination of metal-rich stars at the metal-poor end, and a correction to account for the preferential selection of metal-poor stars by the colour cut used to select the distance limited halo sample. At low galactic latitudes ($20^{\circ} < |b| < 30^{\circ}$) and in the direction of the Galactic anticenter, we identified a metal-rich population at the disc-halo interface, which we concluded to correspond to the Monoceros/TriAnd/ACS/EBS/A13 structure. We also selected a region containing the Virgo overdensity but found no difference in the MDF of this region when compared to other regions of equal Galactic latitude. Comparing the MDF from the current work with others derived in the literature, we find a larger fraction of stars at $\feh < -2$ than in previously published halo MDFs. In particular, we found a slope of $\Delta$(LogN)/$\Delta\feh = 1.0 \pm 0.1$ in the metallicity range of $-3.4 < \feh < -2.5$, which is substantially shallower than previous observational MDFs from spectroscopic samples, but agrees well with the simple closed-box chemical enrichment model of \citet{Hartwick76}, although the model overestimates the number of stars at $\feh < -3.5$. We compare the MDF to that provided in \citet{Helmi18} and \citet{Gallart19} for the Gaia Enceladus structure, but despite some qualitative similarities we cannot conclude whether or not the halo MDF we present is consistent with a single large merger event dominating the stellar content in the inner halo. We see no clear evidence of duality in the halo MDF, but conclude that we need better distances and cleaner samples of distant giants to probe the inner and outer halo in order to properly address this. Finally, we compare the MDF for Milky Way GCs and the MDF for the halo field stars presented in this work, and find that if these two populations follow the same metallicity distribution, the expected number of observed GCs is 10 at $\feh < -2.5$ and 2.8 at $\feh < -3$. We therefore suggest that the lack of GCs at $\feh < -2.5$ in the Milky Way is due to physical processes, rather than statistical undersampling.
\section*{Acknowledgements}

We would like to thank S\o ren Larsen and Mark Gieles for useful discussions which motivated the development of this work, as well as Johannes Schuster for his help in improving the manuscript. We gratefully thank the CFHT staff for performing the observations in queue mode, for their reactivity in adapting the schedule, and for answering our questions during the data-reduction process. We thank Nina Hernitschek for granting us access to the catalogue of Pan-STARRS variability catalogue. 

KY, ES and AA gratefully acknowledge funding by the Emmy Noether program from the Deutsche Forschungsgemeinschaft (DFG). This work has been published under the framework of the IdEx Unistra and benefits from a funding from the state managed by the French National Research Agency as part of the investments for the future program. NFM, gratefully acknowledges support from the French National Research Agency (ANR) funded project ``Pristine'' (ANR-18-CE31-0017) along with funding from CNRS/INSU through the Programme National Galaxies et Cosmologie and through the CNRS grant PICS07708. KY, ES, AA, PJ, VH and NFM benefited from the International Space Science Institute (ISSI) in Bern, CH, thanks to the funding of the Teams "The Formation and Evolution of the Galactic Halo" and "Pristine". JIGH acknowledges financial support from the Spanish Ministry of Science, Innovation and Universities (MICIU) under the 2013 Ram\'on y Cajal program MICIU RYC-2013-14875, and also from the Spanish ministry project MICIU AYA2017-86389-P. CL thanks the Swiss National Science Foundation for supporting this research through the Ambizione grant number PZ00P2\_168065.

Based on observations obtained with MegaPrime/MegaCam, a joint project of CFHT and CEA/DAPNIA, at the Canada-France-Hawaii Telescope (CFHT) which is operated by the National Research Council (NRC) of Canada, the Institut National des Sciences de l'Univers of the Centre National de la Recherche Scientifique of France, and the University of Hawaii. SDSS-IV is managed by the Astrophysical Research Consortium for the 
The Pan-STARRS1 Surveys (PS1) have been made possible through contributions of the Institute for Astronomy, the University of Hawaii, the Pan-STARRS Project Office, the Max-Planck Society and its participating institutes, the Max Planck Institute for Astronomy, Heidelberg and the Max Planck Institute for Extraterrestrial Physics, Garching, The Johns Hopkins University, Durham University, the University of Edinburgh, Queen's University Belfast, the Harvard-Smithsonian Center for Astrophysics, the Las Cumbres Observatory Global Telescope Network Incorporated, the National Central University of Taiwan, the Space Telescope Science Institute, the National Aeronautics and Space Administration under Grant No. NNX08AR22G issued through the Planetary Science Division of the NASA Science Mission Directorate, the National Science Foundation under Grant No. AST-1238877, the University of Maryland, and Eotvos Lorand University (ELTE).

Funding for the Sloan Digital Sky Survey IV has been provided by the Alfred P. Sloan Foundation, the U.S. Department of Energy Office of Science, and the Participating Institutions. SDSS-IV acknowledges support and resources from the Center for High-Performance Computing at the University of Utah. The SDSS web site is www.sdss.org. SDSS-IV is managed by the Astrophysical Research Consortium for the Participating Institutions of the SDSS Collaboration including the Brazilian Participation Group, the Carnegie Institution for Science, Carnegie Mellon University, the Chilean Participation Group, the French Participation Group, Harvard-Smithsonian Center for Astrophysics, Instituto de Astrof\'isica de Canarias, The Johns Hopkins University, Kavli Institute for the Physics and Mathematics of the Universe (IPMU) / University of Tokyo, Lawrence Berkeley National Laboratory, Leibniz Institut f\"ur Astrophysik Potsdam (AIP), Max-Planck-Institut f\"ur Astronomie (MPIA Heidelberg), Max-Planck-Institut f\"ur Astrophysik (MPA Garching), Max-Planck-Institut f\"ur Extraterrestrische Physik (MPE), National Astronomical Observatories of China, New Mexico State University, New York University, University of Notre Dame, Observat\'ario Nacional / MCTI, The Ohio State University, Pennsylvania State University, Shanghai Astronomical Observatory, United Kingdom Participation Group,Universidad Nacional Aut\'onoma de M\'exico, University of Arizona, University of Colorado Boulder, University of Oxford, University of Portsmouth, University of Utah, University of Virginia, University of Washington, University of Wisconsin, Vanderbilt University, and Yale University.




\bibliographystyle{mnras}
\bibliography{bibliography} 




\appendix

\section{Tables of the MDFs and corrections}
\label{appendix}

\begin{table}
\centering
\label{mdf_table}
\caption{MDFs for the raw \pris halo sample, the sample corrected with the GMM, and the sample corrected with the GMM and colour cuts, plotted in Figure \ref{mdf_lit_comp_log} as the grey, cyan, and black histograms, respectively.}
\begin{tabular}{cccc}
\textbf{{[}Fe/H{]}} & \textbf{raw \emph{Pristine}} & \textbf{GMM} & \textbf{corrected} \\ \hline
-3.95               & 395               & 10           & 10                 \\
-3.85               & 143               & 34           & 34                 \\
-3.75               & 170               & 59           & 59                 \\
-3.65               & 294               & 104          & 104                \\
-3.55               & 500               & 149          & 149                \\
-3.45               & 657               & 192          & 192                \\
-3.35               & 787               & 353          & 353                \\
-3.25               & 1037              & 466          & 466                \\
-3.15               & 1189              & 625          & 628                \\
-3.05               & 1458              & 765          & 767                \\
-2.95               & 1650              & 984          & 989                \\
-2.85               & 1999              & 1241         & 1255               \\
-2.75               & 2235              & 1491         & 1520               \\
-2.65               & 2516              & 1808         & 1847               \\
-2.55               & 2845              & 2098         & 2154               \\
-2.45               & 3186              & 2531         & 2634               \\
-2.35               & 3521              & 2816         & 2996               \\
-2.25               & 3770              & 3142         & 3366               \\
-2.15               & 4116              & 3596         & 3892               \\
-2.05               & 4503              & 3910         & 4342               \\
-1.95               & 4635              & 4486         & 5174               \\
-1.85               & 4778              & 4730         & 5680               \\
-1.75               & 4798              & 4951         & 6178               \\
-1.65               & 4650              & 5054         & 6557               \\
-1.55               & 4409              & 5086         & 7092               \\ \hline
-1.45               & 4011              & 4925         & 7589               \\
-1.35               & 3415              & 4529         & 7841               \\
-1.25               & 2865              & 4124         & 7889               \\
-1.15               & 2259              & 3588         & 8787               \\
-1.05               & 1816              & 3101         & 12421              \\
-0.95               & 1326              & 2477         & -                  \\
-0.85               & 1051              & 1875         & -                  \\
-0.75               & 763               & 1426         & -                  \\
-0.65               & 538               & 997          & -                  \\
-0.55               & 415               & 779          & -                  \\
-0.45               & 310               & 483          & -                  \\
-0.35               & 264               & 331          & -                  \\
-0.25               & 217               & 200          & -                  \\
-0.15               & 218               & 130          & -                  \\
-0.05               & -                 & 63           & -                 
\end{tabular}
\end{table}

\begin{table}
\centering
\label{corr_frac}
\caption{Correction factor applied for the colour cut to select the halo sample. The rightmost columns is the correction factor applied to produce the black histogram in Figure \ref{mdf_lit_comp_log} and is the mean of the 10 and 12 Gyr columns.}
\begin{tabular}{cccccc}
\textbf{{[}Fe/H{]}} & \textbf{8 Gyr} & \textbf{10 Gyr} & \textbf{12 Gyr} & \textbf{14 Gyr} & \textbf{Correction} \\ \hline
-3.95               & 1.000          & 1.000           & 1.000           & 1.000           & 1.000               \\
-3.85               & 1.000          & 1.000           & 1.000           & 1.000           & 1.000               \\
-3.75               & 1.000          & 1.000           & 1.000           & 1.000           & 1.000               \\
-3.65               & 1.000          & 1.000           & 1.000           & 1.000           & 1.000               \\
-3.55               & 1.000          & 1.000           & 1.000           & 1.000           & 1.000               \\
-3.45               & 1.000          & 1.000           & 1.000           & 1.000           & 1.000               \\
-3.35               & 1.000          & 1.000           & 1.000           & 1.000           & 1.000               \\
-3.25               & 1.000          & 1.000           & 1.000           & 1.000           & 1.000               \\
-3.15               & 1.009          & 1.000           & 1.011           & 1.031           & 1.006               \\
-3.05               & 1.007          & 1.003           & 1.003           & 1.027           & 1.003               \\
-2.95               & 1.006          & 1.012           & 1.000           & 1.011           & 1.006               \\
-2.85               & 1.009          & 1.009           & 1.015           & 1.018           & 1.012               \\
-2.75               & 1.000          & 1.016           & 1.024           & 1.006           & 1.020               \\
-2.65               & 1.005          & 1.007           & 1.037           & 1.000           & 1.022               \\
-2.55               & 1.007          & 1.017           & 1.037           & 1.014           & 1.027               \\
-2.45               & 1.010          & 1.034           & 1.048           & 1.041           & 1.041               \\
-2.35               & 1.012          & 1.049           & 1.079           & 1.059           & 1.064               \\
-2.25               & 1.024          & 1.058           & 1.085           & 1.112           & 1.072               \\
-2.15               & 1.050          & 1.069           & 1.096           & 1.141           & 1.083               \\
-2.05               & 1.083          & 1.090           & 1.131           & 1.191           & 1.111               \\
-1.95               & 1.111          & 1.127           & 1.180           & 1.258           & 1.154               \\
-1.85               & 1.129          & 1.160           & 1.242           & 1.330           & 1.201               \\
-1.75               & 1.156          & 1.193           & 1.303           & 1.429           & 1.248               \\
-1.65               & 1.212          & 1.239           & 1.356           & 1.594           & 1.298               \\
-1.55               & 1.267          & 1.297           & 1.492           & 1.835           & 1.395               \\ \hline
-1.45               & 1.315          & 1.403           & 1.679           & 2.054           & 1.541               \\
-1.35               & 1.391          & 1.531           & 1.932           & 2.338           & 1.732               \\
-1.25               & 1.479          & 1.689           & 2.137           & 5.174           & 1.913               \\
-1.15               & 1.635          & 2.029           & 2.869           & -               & 2.449               \\
-1.05               & 1.841          & 2.411           & 5.600           & -               & 4.006               \\
-0.95               & 2.209          & 2.658           & -               & -               & -                   \\
-0.85               & 2.471          & 10.235          & -               & -               & -                   \\
-0.75               & 3.101          & -               & -               & -               & -                   \\
-0.65               & 20.823         & -               & -               & -               & -                   \\
-0.55               & -              & -               & -               & -               & -                   \\
-0.45               & -              & -               & -               & -               & -                   \\
-0.35               & -              & -               & -               & -               & -                   \\
-0.25               & -              & -               & -               & -               & -                   \\
-0.15               & -              & -               & -               & -               & -                   \\
-0.05               & -              & -               & -               & -               & -                  
\end{tabular}
\end{table}

\begin{table}
\centering
\label{gmm_params}
\caption{Best fit parameters for the GMM shown in the right panel of Figure \ref{gmm}.}
\begin{tabular}{cccc}
\textbf{Component} & \textbf{Mean} & \textbf{Weight} & \textbf{Covariance matrix} \\ 
& \textbf{(x,y)} & & $\begin{bmatrix}
\sigma^2_x & \rho\sigma_x\sigma_y \\
\rho\sigma_x\sigma_y & \sigma^2_x
\end{bmatrix} $ \\

\hline 
\textbf{1}                                                     & (-1.320, -1.501)                                                 & 0.28                                                           & $\begin{bmatrix}
0.183 & 0.113 \\
0.113 & 0.188
\end{bmatrix} $                                                        \\ \\
\textbf{2}                                                     & (-1.893, -2.121)                                                 & 0.39                                                           & $\begin{bmatrix}
0.232 & 0.197 \\
0.197 & 0.222
\end{bmatrix} $                                                             \\ \\
\textbf{3}                                                     & (-2.448, -2.815)                                                 & 0.33                                                           & $\begin{bmatrix}
0.251 & 0.033 \\
0.033 & 0.127
\end{bmatrix} $                                                            
\end{tabular}
\end{table}


\bsp	
\label{lastpage}
\end{document}